\documentclass[10pt,conference,compsocconf]{IEEEtran}
\usepackage{graphicx}
\usepackage{subfig}
\usepackage{amsmath}
\usepackage{amssymb}
\usepackage{amsfonts,dsfont}
\usepackage{times}
\usepackage{cite}
\usepackage{caption}
\usepackage{verbatim}
\usepackage{algorithm}
\usepackage{algorithmic}
\usepackage{hyperref}
\usepackage{breakurl}
\usepackage{url}
\usepackage{cite}
\usepackage{color}

\newcommand{\bpi}{\boldsymbol{\pi}}

\newcommand{\bQ}{\boldsymbol{Q}}
\newcommand{\bP}{\boldsymbol{P}}
\newcommand{\bD}{\boldsymbol{D}}
\newcommand{\bI}{\boldsymbol{I}}
\newcommand{\bv}{\boldsymbol{v}}

\newtheorem{theorem}{{\bf \hspace{-0.18in} Theorem}}

\def\done{\hspace*{\fill} \rule{1.8mm}{2.5mm} }

\IEEEoverridecommandlockouts

\begin{document}

\title{Stochastic Analysis on RAID Reliability for Solid-State Drives
\vspace{-12pt}}

\author{
\IEEEauthorblockN{Yongkun Li, Patrick P. C. Lee, John C. S. Lui}
\IEEEauthorblockA{The Chinese University of Hong Kong\\
 Email:yongkunlee@gmail.com, \{pclee,cslui\}@cse.cuhk.edu.hk}
 }

\maketitle

\begin{abstract}
Solid-state drives (SSDs) have been widely deployed in desktops and data
centers.  However, SSDs suffer from bit errors, and the bit error rate is
 {\em time dependent} since it increases as an SSD wears down.
Traditional storage systems mainly use parity-based RAID to provide
reliability guarantees by striping redundancy across multiple devices, but
the effectiveness of RAID  in SSDs remains debatable as parity updates
aggravate the wearing and bit error rates of SSDs. In particular, an
 open problem is that how different parity distributions over multiple
devices, such as the even distribution suggested by conventional wisdom,
or uneven distributions proposed in recent RAID schemes for SSDs, may
influence the reliability of an SSD RAID array. To address this
fundamental problem, we propose the first analytical model to quantify the
reliability dynamics of an SSD RAID array. Specifically, we develop a
``non-homogeneous'' continuous time Markov chain model, and derive the
transient reliability solution. We validate our model via trace-driven
simulations and conduct numerical analysis to provide insights into the
reliability dynamics of SSD RAID arrays under different parity
distributions and  subject to different bit error rates and array
configurations. Designers can use our model to decide the appropriate
parity distribution based on their reliability requirements.
\end{abstract}


\section{{\bf Introduction}}
\label{sec: introduction}

Solid-state drives (SSDs) emerge to be the next-generation storage medium.
Today's SSDs mostly build on NAND flash memories, and provide several
 design enhancements over  hard disks including higher  I/O
performance, lower energy consumption, and higher shock resistance.  As SSDs
continue to see price drops nowadays, they have been widely deployed in
desktops and large-scale data centers
\cite{SSDinDataCenter2,SSDinDataCenter3}.

However, even though enterprise SSDs generally provide high reliability
guarantees (e.g., with mean-time-between-failures of 2~million hours
\cite{intelspec}), they are susceptible to wear-outs and bit errors. First,
SSDs regularly perform erase operations between writes, yet they can only
tolerate a limited number of erase cycles before wearing out. For example,
the erasure limit is only 10K for multi-level cell (MLC) SSDs \cite{Chen09},
and even drops to several hundred for the latest triple-level cell (TLC)
SSDs \cite{Grupp12}. Also, bit errors are common in SSDs due to read
disturbs, program disturbs, and retention errors
\cite{Mielke08,Grupp09,Grupp12}.  Although in practice SSDs use error
correction codes (ECCs) to protect data \cite{Deal09, Mariano12}, the
protection is limited since the bit error rate increases as SSDs issue more
erase operations \cite{Mielke08,Grupp09}. We call a post-ECC bit error an
{\em uncorrectable bit error}. Furthermore, bit errors become more severe
when the density of flash cells increases and the feature size decreases
\cite{Grupp12}.  Thus, SSD reliability remains a legitimate concern,
especially when an SSD issues frequent erase operations due to heavy writes.

RAID (redundant array of independent disks) \cite{Patterson88} provides an
option to improve reliability of SSDs.  Using parity-based RAID (e.g.,
RAID-4, RAID-5), the original data is encoded into {\em parities}, and the
data and parities are striped across multiple SSDs to provide storage
redundancy against failures.  RAID has been widely used in tolerating hard
disk failures, and conventional wisdom suggests that parities should be {\em
evenly} distributed across multiple drives so as to achieve better load
balancing, e.g., RAID-5. However, traditional RAID introduces a different
reliability problem to SSDs since  parities are updated for every data write
and this  aggravates the erase cycles. To address this problem, authors in
\cite{Balakrishnan10} propose a  RAID scheme called {\em Diff-RAID} which
aims to enhance the SSD RAID reliability  by keeping {\em uneven} parity
distributions. Other studies (e.g., \cite{Im11, Kim12, Lee11, Lee09, Park09,
Mao12}) also explore the use of RAID in SSDs.

However, there remain open issues on the proper architecture designs of
highly reliable SSD RAID \cite{Jeremic11}.  One specific open problem is how
different parity distributions generally influence the reliability of an SSD
RAID array subject to different error rates and array configurations. {\em
In other words, should we distribute parities evenly or unevenly across
multiple SSDs with respect to the SSD RAID reliability?} This motivates us
to characterize the SSD RAID reliability using analytical modeling, which
enables us to readily tune different input parameters and determine their
impacts on reliability.  However, analyzing the SSD RAID reliability is
challenging, as the error rates of SSDs are {\em time-varying}.
Specifically, unlike hard disk drives in which error arrivals are commonly
modeled as a constant-rate Poisson process (e.g., see \cite{Schulze89,
Muntz90}), SSDs have an increasing error arrival rate as they wear down with
more erase operations.

In this paper, we formulate a continuous time Markov chain (CTMC) model to
analyze the effects of different parity placement strategies, such as
traditional RAID-5 and Diff-RAID \cite{Balakrishnan10}, on the reliability
dynamics of an SSD RAID array.  To capture the time-varying bit error rates
in SSDs, we formulate a {\em non-homogeneous} CTMC model, and conduct
transient analysis to derive the system reliability at any specific time
instant.  To our knowledge, this is the {\em first} analytical study on the
reliability of an SSD RAID array.

In summary, this paper makes two key contributions:
\begin{itemize}
\item We formulate a non-homogeneous CTMC model to characterize the
    reliability dynamics of an SSD RAID array. We use the uniformization
    technique \cite{de00,Jensen53,Reibman89} to derive the transient
    reliability of the array.  Since the state space of our model
    increases with the SSD size, we develop optimization techniques to
    reduce the computational cost of transient analysis. We also quantify
    the corresponding error bounds of the uniformization and optimization
    techniques.  Using the SSD simulator \cite{Agrawal08}, we validate our
    model via trace-driven simulations.
\item We conduct extensive numerical analysis to compare the reliability
    of an SSD RAID array under RAID-5 and Diff-RAID \cite{Balakrishnan10}.
    We observe that Diff-RAID, which places parities unevenly across SSDs,
    only improves the reliability over RAID-5 when the error rate is not
    too large, while RAID-5 is reliable enough if the error rate is
    sufficiently small.  On the other hand, when the error rate is very
    large, neither RAID-5 nor Diff-RAID can provide high reliability, so
    increasing fault tolerance (e.g., RAID-6 or a stronger ECC) becomes
    necessary.
\end{itemize}

The rest of this paper proceeds as follows. In Section~\ref{sec: model}, we
formulate our model that characterizes the reliability dynamics of an SSD
RAID array, and formally define the reliability metric. In Section~\ref{sec:
transient}, we derive the transient system state using uniformization and
some optimization techiniques. In Section~\ref{sec: validation}, we validate
our model via trace-driven simulations. In Section~\ref{sec: simulation}, we
present numerical analysis results on how different parity placement
strategies influence the RAID reliability. Section~\ref{sec: related}
reviews related work, and finally Section~\ref{sec: conclusion} concludes.

\section{{\bf System Model}}
\label{sec: model}

It is well known that RAID-5 is effective in providing single-fault
tolerance for traditional hard disk storage. It distributes parities evenly
across all drives and achieves load balancing.  Recently, Balakrishnan {\em
et al.} \cite{Balakrishnan10} report that RAID-5 may result in correlated
failures, and hence poor reliability, for SSD RAID arrays if SSDs are worn
out at the same time.  Thus, they propose a modified RAID scheme called {\em
Diff-RAID} for SSDs.  Diff-RAID improves RAID-5 through (i) distributing
parties unevenly and (ii) redistributing parities each time when a worn-out
SSD is replaced so that the oldest SSD always has the most parities and
wears out first. However, it remains unclear whether Diff-RAID (or placing
parities unevenly across drives) really improves the reliability of SSD RAID
over RAID-5 in all error patterns, as there is a lack of comprehensive
studies on the reliability dynamics of SSD RAID arrays under different
parity distributions.

In this section, we first formulate an SSD RAID array, then characterize the
age of each SSD based on the age of the array (we will formally define the
concept of age in later part of this section). Lastly, we model the error
 rate based on the age of each SSD, and formulate a {\em
non-homogeneous CTMC}  to characterize the reliability dynamics of an SSD
RAID array under various parity distributions, including different parity
placement distributions like RAID-5 or Diff-RAID.
Table~\ref{table:notations} lists the major notations used in this paper.

\begin{table}[!h]
\begin{center}
{\small
\begin{tabular*}{\columnwidth}{p{0.04\columnwidth} p{0.0001\columnwidth} p{0.84\columnwidth}}
  \hline
  \hline
  \multicolumn{3}{l}{ {\bf Specific Notations of SSD}}    \\
  \hline
  $M$ & : & Erasure limit of each block (e.g., 10K)\\
  $B$ & : & Total number of blocks in each SSD\\
  $\lambda_i(t)$ & :&  Error rate of a chunk in SSD~$i$ at time $t$\\
  \hline
  \multicolumn{3}{l}{{\bf Specific Notations of RAID Array}}    \\
  \hline
  $N$ & :&  Number of data drives (i.e., an array has $N+1$ SSDs)\\
  $S$ & :&  Total number of stripes in an SSD RAID array\\
  $p_i$ & :& Fraction of parity chunks in SSD~$i$, and $\sum_{i=0}^{N}p_i=1$\\
  $k$ & :&  Total number of erasures performed on SSD RAID array (i.e., system age of the array)\\
  $k_i$ & :& Number of erasures performed on each block of SSD~$i$ (i.e., age of SSD~$i$)\\
  $T$   & :& Average inter-arrival time of two consecutive erasure operations on SSD RAID array   \\
  $\pi_j\!(t)$ & :& Probability that the array has $j$ stripes that contain
                  exactly one erroneous chunk each, ($0\leq j\leq S$)\\
  $\!\!\pi\!_{S\!+\!1}\!(t)$ & :& Probability that at least one stripe of the array contains
                  more than one erroneous chunk, so $\sum_{j=0}^{S+1}\pi_j(t)=1$\\
  $R(t)$ & :& Reliability at time $t$, i.e., probability that no data loss happens until time $t$,
              $R(t)=\sum_{j=0}^{S}\pi_j(t)$\\
  \hline\hline
\end{tabular*}
}
\end{center}
\caption{Notations.}\label{table:notations}
\end{table}

\subsection{SSD RAID Formulations}

An SSD  is usually organized in {\em blocks}, each of which typically
contains 64 or 128 {\em pages}. Both read and program (write) operations are
performed in unit of pages, and  each page is of size 4KB.  Data can only be
programmed to {\em clean} pages.  SSDs use an {\em erase} operation, which
is performed in unit of blocks, to reset all pages in a block into clean
pages.  To improve   write performance, SSDs use {\em out-of-place} writes,
i.e., to update a page, the new data is programmed to a clean page while the
original page is marked as invalid.  An SSD is usually composed of multiple
{\em chips} (or packages), each containing thousands of blocks. Chips are
independent of each other and can operate in parallel.  We refer readers to
\cite{Agrawal08} for a detailed description about the SSD organization.

\begin{figure}[t]
  \centering
  \includegraphics[width=6cm]{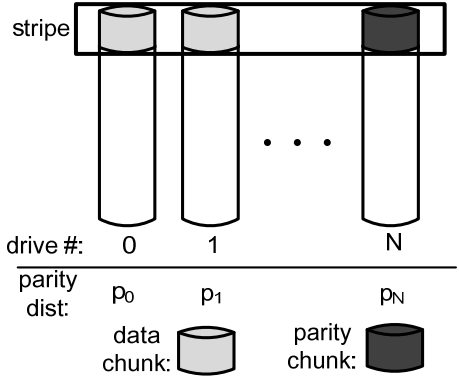}\\
  \caption{Organization of an SSD RAID array.}
\label{fig: raid_structure}
\end{figure}

We now describe the organization of an SSD RAID array that we consider, as
shown in Figure~\ref{fig: raid_structure}.  We consider the device-level
RAID organization where the array is composed of $N\!+\!1$ SSDs numbered
from 0 to $N$.  In this paper, we address the case where the array is
tolerable against a single SSD failure, as assumed in traditional RAID-4,
RAID-5 schemes and the modified RAID schemes for SSDs \cite{Balakrishnan10,
Im11, Lee11, Lee09, Park09, Kim12, Mao12}. Each SSD is divided into multiple
non-overlapping {\em chunks}, each of which can be mapped to one or multiple
physical pages. The array is further divided into {\em stripes}, each of
which is a collection of $N+1$ chunks from the $N+1$ SSDs.  Within a stripe,
there are $N$ data chunks, and one parity chunk encoded from the $N$ data
chunks.  We call a chunk an {\em erroneous chunk} when  uncorrectable bit
errors appear in that chunk; or a {\em correct chunk} otherwise. Since we
focus on single-fault tolerance, we require that each stripe contains at
most one erroneous chunk without data loss so that it can be recovered from
other surviving chunks in the same stripe.

Suppose that each SSD contains $B$ blocks, and the array contains $S$
stripes (i.e., $S$ chunks per SSD). For simplicity, we assume that all $S$
stripes are used for data storage.
To generalize our analysis, we organize parity chunks in the array according
to some probability distribution. We let \mbox{SSD $i$} contain a fraction
$p_i$ of parity chunks. In the special case of RAID-5, parity chunks are
{\em evenly placed}  across all devices, so $p_i = \frac{1}{N+1}$ for all
$i$ if the array consists of $N+1$ drives.  For Diff-RAID, $p_i$'s do not
need to be equal to $\frac{1}{N+1}$, but only need to satisfy the condition
of $\sum_{i=0}^N p_i = 1$.

Each block in an SSD can only sustain a limited number of erase cycles, and
is supposed to be worn out after the limit.  We denote the erasure limit by
$M$, which corresponds to the lifetime of a block.  To enhance the
durability of SSDs, efficient wear-leveling techniques are often used to
balance the number of erasures across all blocks.  In this paper, we assume
that each SSD achieves perfect wear-leveling such that every block has
exactly the same number of erasures.  Let $k_i$ ($\le M$) be the number of
erasures that have been performed on each block in SSD $i$, where $0\le i\le
N$.  We denote $k_i$ as the {\em age} of each block in SSD $i$, or
equivalently, the age of SSD $i$ when perfect wear-leveling is assumed. When
an SSD reaches its erasure limit, we assume that it is replaced by a new
SSD. For simplicity, we treat $k_i$ as a continuous value in $[0,M]$. Let
$k$ be the total number of erase operations that the whole array has
processed, and we call $k$ the {\em system age} of the array.

\subsection{SSD Age Characterization}

In this subsection, we proceed to characterize the  age of each SSD for a
given RAID scheme. In particular, we derive $k_i$, denoting the age of SSD
$i$, when the whole array has already performed a total of $k$ erase
operations. This characterization enables us to model the error rate in each
SSD accurately (see Section~\ref{subsec:ctmc}). We focus on two RAID
schemes: traditional RAID and Diff-RAID \cite{Balakrishnan10}.

We first quantify the {\em aging rate} of each SSD in an array. Let $r_i$ be
the aging rate of SSD $i$.  Note that for each stripe, updating a data chunk
also has the parity chunk updated.  Suppose that each data chunk has the
same probability of being accessed.  On average, the {\em ratio} of the
aging rate of SSD $i$ to that of SSD $j$ can be expressed as
\cite{Balakrishnan10}:
\begin{equation}
  \frac{r_i}{r_j}=\frac{p_iN+(1-p_i)}{p_jN+(1-p_j)}.
  \label{eq: aging_ratio}
\end{equation}
Equation~(\ref{eq: aging_ratio}) states that the parity chunk ages $N$ times
faster than each data chunk.  Given the aging rates $r_i$'s, we can quantify
the probability of SSD~$i$ being the target drive for each erase operation,
which we denote by $q_i$. We model $q_i$ by making it proportional to the
aging rate of SSD $i$, i.e.,
\begin{equation}
  q_i = \frac{r_i}{\sum_{i=0}^Nr_i}=\frac{p_iN+(1-p_i)}{\sum_{i=0}^N(p_iN+(1-p_i))}.
  \label{eq: erase_prob}
\end{equation}

We now  characterize the age of Diff-RAID which places parities unevenly and
redistributes  parity chunks after the worn-out SSD is replaced so as to
maintain the age ratios and always wear out the oldest SSD first. To
mathematically characterize the system age of Diff-RAID, define $A_i$ as the
remaining fraction of erasures that SSD $i$ can sustain right after an SSD
replacement. Clearly, $A_i=1$ for a brand-new drive and $A_i=0$ for a
worn-out drive. Without loss of generality, we assume that the drives are
sorted by $A_i$ in descending order, i.e., $A_0\geq A_1\geq\cdots\geq A_N$,
and we have $A_0=1$ as it is the newly replaced drive.  Diff-RAID performs
parity redistribution to guarantee that the aging ratio in
Equation~(\ref{eq: aging_ratio}) remains unchanged. Therefore, the remaining
fraction of erasures for each drive will {\em converge}, and the values of
$A_i$'s in the steady state are given by \cite{Balakrishnan10}:
\begin{equation}
 A_i \!=\! \frac{\sum_{j=i}^Nr_j}{\sum_{j=0}^Nr_j} \!=\!
  \frac{\sum_{j=i}^N(p_jN\!+\!(1\!-\!p_j))}{\sum_{j=0}^N(p_jN\!+\!(1\!-\!p_j))},
  \quad 0\!\leq\! i\!\leq\! N.
\label{eq: age_dist_convergent}
\end{equation}
In this paper, we study Diff-RAID after the age distribution of SSDs right
after each drive replacement converges, i.e., the initial remaining
fractions of erasures of SSDs in Diff-RAID follow the distribution of
$A_i$'s in Equation~(\ref{eq: age_dist_convergent}).

 We now characterize
$k_i$ for Diff-RAID. Recall that each SSD has $B$ blocks. Due to perfect
wear-leveling, every block of SSD $i$ has the same probability $q_i / B$ of
being the target block for an erase operation. Thus, if the array has
processed $k$ erase operations, the age of SSD $i$ is:
\begin{equation}
 \mbox{{\bf Diff-RAID:}}
\quad k_i=\Big(\frac{kq_i}{B} \ \mbox{\sf mod} \ \frac{q_i}{q_N}(M\!-\!k_{N0})\Big)
      \!+\! k_{i0},
\label{eq: D-RAIDblock_age1}
\end{equation}
where $k_{i0} = M(1\!-\!A_i)$ is the initial number of times that each block
of SSD $i$ has been erased right after a drive replacement, and the notation
{\sf mod} denotes the modulo operation. The rationale of Equation~(\ref{eq:
D-RAIDblock_age1}) is as follows. Since we sort the SSDs by $A_i$ in
descending order, SSD~$N$ always has the highest aging rate and will be
replaced first.  Thus, after each block of SSD~$N$ has performed
$(M\!-\!k_{N0})$ erasures, SSD~$N$ will be replaced, and each block of
SSD~$i$ has just been erased $\frac{q_i}{q_N}(M\!-\!k_{N0})$ times.
Therefore, for SSD~$i$, a drive replacement happens when each block has been
erased {\em every} $\frac{q_i}{q_N}(M-k_{N0})$ times.  Moreover, the initial
number of erasures on each block of SSD~$i$  right after a drive replacement
is $k_{i0}$. Thus, the age of SSD~$i$ is derived as in Equation~(\ref{eq:
D-RAIDblock_age1}). Since $k_{i0}=M(1-A_i)$ and $A_N=q_N$,
Equation~(\ref{eq: D-RAIDblock_age1}) can be rewritten as:
\begin{equation}
 \mbox{{\bf Diff-RAID:}} \quad k_i=\left((kq_i / B) \ \mbox{\sf mod} \ Mq_i\right) +
 M(1-A_i).
\label{eq: D-RAIDblock_age2}
\end{equation}

For traditional RAID (e.g., RAID-4 or RAID-5), parity chunks are kept
intact, and will not be redistributed after a drive replacement. So after
the array has performed $k$ erase operations, each block of SSD~$i$ has just
performed $kq_i/B$ erasures, and an SSD will be replaced every time when
each block performed $M$ erasures.  Thus, the age of SSD~$i$ is:
\begin{equation}
\mbox{{\bf Traditional RAID:}} \quad k_i =(kq_i/B) \ \mbox{\sf mod} \ M.
\label{eq: T-RAIDblock_age}
\end{equation}

\subsection{Continuous Time Markov Chain (CTMC)}
\label{subsec:ctmc}

We first model the error rate of an SSD. We assume that the error arrival
processes of different chunks in an SSD are independent. Since different
chunks in an SSD have the same age, they must have the same error rate.  We
let $\lambda_i(t)$ represent the error rate of each chunk in SSD~$i$ at time
$t$, and model it as a function of the number of erasures on SSD~$i$ at time
$t$, which is denoted by $k_i(t)$ (the notation $t$ may be dropped if the
context is clear). Furthermore, to reflect that bit errors increase with the
 number of erasures, we model the error rate based on a Weibull
distribution \cite{Weibull51}, which has been widely used in reliability
engineering. Formally,
\begin{equation}
   \lambda_i(t) = c\alpha(k_i(t))^{\alpha-1}, \quad \alpha \!> \!1,
  \label{eq: error_rate}
\end{equation}
where $\alpha$ is called the {\em shape parameter} and $c$ is a constant.

Note that even if the error rates of SSDs are time-varying, they only vary
with the number of erasures on the SSDs. If we let $t_k$ be the time point
of the $k^{th}$ erasure on the array, then during the period $(t_k,t_{k+1})$
(i.e., between the $k^{th}$ and $(k+1)^{th}$ erasures), the number of
erasures on each SSD is fixed, hence the error rates during this period
should be constant, and the error arrivals can be modeled as a Poisson
process. In particular, $k_i(t)=k_i(k)$ if $t\in (t_k,t_{k+1})$, and the
function $k_i(k)$ is expressed by Equation~(\ref{eq: D-RAIDblock_age2}) and
(\ref{eq: T-RAIDblock_age}).

We now formulate a CTMC model to characterize the reliability dynamics of an
SSD RAID array.  Recall that the array provides single-fault tolerance for
each stripe.  We say that the CTMC is at state $i$ if and only if the array
has $i$ stripes that contain exactly one erroneous chunk each, where $0 \!
\le \! i\! \le \!S$.  Data loss happens if any one stripe contains more than
one erroneous chunk, and we denote this state by $S\!+\!1$.  Let $X(t)$ be
the system state at time $t$. Formally, we have
  $X(t)\in \{0,1,...,S+1\}, \forall t\geq 0.$ To derive the system state, we let $\pi_j(t)$ be the
probability that the CTMC is at state $j$ at time $t$ ($0 \! \leq \! j \!
\leq \! S \!+ \!1$), so the system state can be characterized by the vector
$\boldsymbol{\pi}(t)=(\pi_0(t),\pi_1(t),...,\pi_{S+1}(t))$.

\begin{figure}[!t]
\centering
\includegraphics[width=6cm]{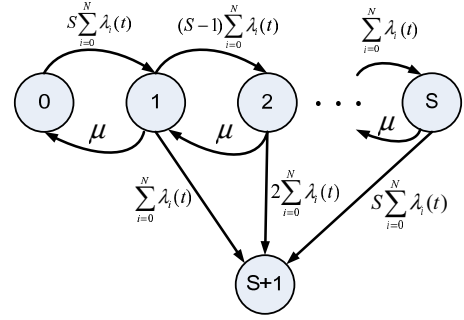}
\caption{State transition  of the non-homogeneous CTMC.}
\label{fig: state}
\end{figure}

Let us consider the transition of the CTMC.  For each stripe, if it contains
one erroneous chunk, then the erroneous chunk can be reconstructed from the
other surviving chunks in the same stripe.  Assume that only one stripe can
be reconstructed at a time, and that the reconstruction time follows an
exponential distribution with rate $\mu$. The state transition diagram  of
the CTMC is depicted in Figure~\ref{fig: state}. To elaborate, suppose that
the RAID array is currently at state $j$, if an erroneous chunk appears in
one of the $(S\!-\!j)$ stripes that originally have no erroneous chunk, then
it will move to state $j\!+\!1$ with rate
$(S\!-\!j)\sum_{i=0}^N\lambda_i(t)$; if an erroneous chunk appears in one of
the $j$ stripes that already have another erroneous chunk, then the system
will move to state $S\!+\!1$ (in which data loss occurs) with rate
$j\sum_{i=0}^N\lambda_i(t)$.

We now define  the {\em reliability} of an SSD RAID array at time $t$, and
denote it by $R(t)$. Formally, it is the probability that no stripe has
encountered data loss until time $t$.
\begin{equation}
R(t) = \sum\nolimits_{j=0}^{S}\pi_j(t).
\label{eq: reliability_def}
\end{equation}

Note that our model captures the time-varying nature of reliability over the
lifespan of the SSD RAID array.  Next, we show how to analyze this
non-homogeneous CTMC.

\section{{\bf Transient Analysis of CTMC}}
\label{sec: transient}

In this section, we derive $\bpi(t)$, the system state of an SSD RAID array
at any time $t$. Once we have $\bpi(t)$, we can then compute the
instantaneous reliability $R(t)$ according to Equation~(\ref{eq:
reliability_def}).  There are two major challenges in deriving $\bpi(t)$.
First, it involves {\em transient analysis}, which is different from the
conventional steady state Markov chain analysis.  Second, the underlying
CTMC $\{X(t), t\geq 0\}$ is {\em non-homogeneous}, as the error arrival rate
$\lambda_i(t)$ is time varying, and it also has a very large state space.

In the following, we first present the mathematical foundation on analyzing
the non-homogeneous CTMC so as to compute the transient system state, then
formalize an algorithm based on the mathematical analysis. At last, we
develop an optimization technique to address the challenge of large state
space of the CTMC so as to further reduce the computational cost of the
algorithm.

\subsection{Mathematical Analysis on the Non-homogeneous CTMC}
\label{subsec: analyze_nonhomogeneous}

Note that the error rates of SSDs within a period $(t_k, t_{k+1})$ ($k=0,
1,2,...$) are  constant, so if we only focus on a particular time period of
the CTMC, i.e., $\{X(t), t_k< t\leq t_{k+1}\}$, then it becomes a
time-homogeneous CTMC. Therefore, the intuitive way to derive the transient
solution of the CTMC $\{X(t), t\geq 0\}$ is to divide it into many
time-homogeneous CTMCs $\{X(t), t_k< t\leq t_{k+1}\}$ $(k=0,1,2...)$, then
use  the {\em uniformization} technique \cite{Jensen53,Reibman89,de00}  to
analyze these time-homogeneous CTMCs one by one in time ascending order.
Specifically, to derive $\bpi(t_{k+1})$, one first derives  $\bpi(t_{1})$
from the initial state $\bpi(0)$, then takes $\bpi(t_{1})$ as the initial
state and derives $\bpi(t_{2})$ from $\bpi(t_{1})$ and so on.

However, this computational approach may take a prohibitively long time to
derive $\bpi(t_{k+1})$ when $k$ is very large, which usually occurs in SSDs.
Since $k$ denotes the number of erasures performed on an SSD RAID array, it
can grow up to $(N\!+\!1)BM$, where both $B$ (the number of blocks in an
SSD) and $M$ (the erasure limit) could be very huge, say, 100K and 10K,
respectively (see Sec.~\ref{sec: simulation}). Therefore, simply applying
the uniformization technique is computationally infeasible to derive the
reliability of an SSD RAID array, especially when the array performs a lot
of erasures.

To overcome the above challenge, we propose an optimization technique which
combines multiple time periods together. The main idea is that since the
difference of the generator matrices at two consecutive periods is very
small in general, we consider $s$ consecutive periods together, where $s$ is
called the {\em step size}. For simplicity of discussion, let $T$ be the
average inter-arrival time of two consecutive erasure operations, i.e.,
$t_k=kT$. To analyze the non-homogeneous CTMC over $s$ periods $\{X(t),
lsT\!<t\!\le(l\!+\!1)sT\}$ ($l=0,1,...$), we define another time-homogeneous
CTMC $\{\tilde{X}(t), lsT\! <\!t\!\le\!(l\!+\!1)sT\}$  to approximate it and
also quantify the error bound. The derivation of $\bpi((l\!+\!1)sT)$ given
$\bpi(lsT)$ proceeds as follows.

\noindent {\bf Step 1: Constructing a time-homogeneous CTMC $\{\tilde{X}(t),
lsT\! <\!t\!\le\!(l\!+\!1)sT\}$ with generator matrix $\tilde{\bQ_l}$.} Note
that there are $s$ periods in the interval $(lsT, (l\!+\!1)sT)$. We denote
the generator matrices of the original Markov chain $\{X(t)\}$ during each
of the $s$ periods  by $\bQ_{ls}$, $\bQ_{ls+1}$, ... , $\bQ_{(l+1)s-1}$. To
construct  $\{\tilde{X}(t), lsT <t\le(l+1)sT\}$, we define $\tilde{\bQ_l}$
as a function of the $s$ generator matrices.
\begin{equation}
  \tilde{\bQ_l} = f(\bQ_{ls},\bQ_{ls+1},..., \bQ_{(l+1)s-1}), \quad
  l=0,1,...
\label{eq: new_generator matrix}
\end{equation}

Intuitively, $\tilde{\bQ_l}$ can be viewed as the ``average'' over the $s$
generator matrices.  To illustrate, consider a special case where $\alpha$
in Equation~(\ref{eq: error_rate}) is set to be $\alpha=2$.  Then the error
arrival rate of each chunk of SSD $i$ becomes $2ck_i$.  In this case, each
element of the generator matrix $\bQ_k$ becomes
\begin{equation}
  q_{i,j}(k)\!\!=\!\!\left\{
            \begin{aligned}
              &-\!S\Sigma, & i \!= j \! = \!0, \\
              &-\!\mu\!-\!S\Sigma, \hspace{-24pt} & 0\!<\! i \!\leq \!S, \ j\!=\!i,\\
              &(S\!-\!i)\Sigma, & 0\!\leq \!i\! < \!S, \ j\!=\!i\!+\!1,\\
              &i\Sigma, & 0\!< \!i \!\leq \!S, \ j\!=\!S\!+\!1,\\
              &\mu, & 0\!< \!i \!\leq\! S, \ j=i\!-\!1,\\
              &0, & \mbox{ otherwise },
            \end{aligned}
            \right.
  \label{eq: generator_matrix_alpha2}
\end{equation}
where $\Sigma=\sum\nolimits_{i=0}^N\!2ck_i$ and $k_i$ is computed by
Equations~(\ref{eq: D-RAIDblock_age2}) and (\ref{eq: T-RAIDblock_age}). Now,
for the Markov chain $\tilde{X}(t)$, we let $\tilde{\bQ_l}$ be an average of
these $s$ generator matrices $\bQ_k$. Mathematically,
\begin{equation}
  \tilde{\bQ_l} = \Big(\sum\nolimits_{k=ls}^{(l+1)s-1}\bQ_k\Big) / s, \quad l=0,1,...
\label{eq: new_generator matrix_avg}
\end{equation}
Note that our analysis is applicable for other values of $\alpha$, with
different choices of defining $\tilde{\bQ_l}$ in Equation~(\ref{eq:
new_generator matrix}) and different error bounds.  We pose the further
analysis of different values of $\alpha$ as future work.  In the following
discussion, we fix $\alpha=2$, whose error bound can be derived.

\noindent {\bf Step 2: Deriving the system state $\tilde{\bpi}((l+1)sT)$
under the time-homogeneous CTMC $\{\tilde{X}(t)\}$.} To derive the system
state at time $(l\!+\!1)sT$, which we denote as $\tilde{\bpi}((l\!+\!1)sT)$,
we solve the Kolmogorov's forward equation and we have
\begin{equation}
  \tilde{\bpi}((l\!+\!1)sT)\!\!=\!\!\tilde{\bpi}(lsT)\!\!\sum\nolimits_{n=0}^{\infty} (\tilde{\bQ_l}sT)^n /n!,
  \ l=0,1,...
\label{eq: system_state_mp}
\end{equation}
where the initial state is $\tilde{\bpi}(0)=\bpi(0)$.

\noindent {\bf Step 3: Applying uniformization to solve Equation~(\ref{eq:
system_state_mp}).}  We let $\tilde{\Lambda}_l\geq \max_{ls \leq k \leq
(l+1)s-1}\max_{0\leq i\leq S+1}|-q_{i,i}(k)|$, and let $\tilde{\bP_l}
=\bI+\frac{\tilde{\bQ_l} }{\tilde{\Lambda}_l}$. Based on the uniformization
technique \cite{de00}, the system state at time $(l+1)sT$ can be derived as
follows.
\begin{equation}
      \tilde{\bpi}((l\!+\!1)sT) \!\!=\!\! \!\sum\nolimits_{n=0}^{\infty}\!e^{-\tilde{\Lambda}_l sT}\frac{(\tilde{\Lambda}_l sT)^n}{n!}\bv_l(n), l\!=\!0, 1, ...
      \label{eq: system_state_mp_u}
\end{equation}
where $\bv_l(n)=\bv_l(n-1)\tilde{\bP}_l$ and $\bv_l(0)=\tilde{\bpi}(lsT)$.
The initial state is $\tilde{\bpi}(0)=\bpi(0)$.

\noindent {\bf Step 4: Truncating the infinite summation in
Equation~(\ref{eq: system_state_mp_u}) with a quantifiable error bound.} We
denote the truncation point for interval $(lsT, (l\!+\!1)sT)$ by $U_l$ and
denote the system state at time $(l\!+\!1)sT$ after truncation by
$\hat{\tilde{\bpi}}((l\!+\!1)sT)$. We also denote the error caused by
combining $s$ periods together and truncating the infinite series in
interval $(lsT, (l\!+\!1)sT)$ by
$\hat{\tilde{\epsilon}}_l||\hat{\tilde{\bpi}}((l+1)sT) - \bpi((l+1)sT)||_1$,
where $\bpi((l\!+\!1)sT)$ denotes the accurate system state obtained by
iteratively analyzing the time-homogeneous CTMCs $\{X(t),kT<t\leq (k+1)T\}$
($k=0,1,...,(l\!+\!1)s-1$) from the initial state $\bpi(0)$. Now,
$\hat{\tilde{\bpi}}((l\!+\!1)sT)$ and $\hat{\tilde{\epsilon}}_l$ can be
computed using the following theorem.

\begin{theorem}
After truncating the infinite series, the system state at time $(l+1)sT$ for
the Markov chain $\{\tilde{X}(t)\}$ with step size $s$ can be computed as
follows.
\begin{equation}
      \hat{\tilde{\bpi}}((l\!+\!1)sT) \!\!=\!\!\! \sum\nolimits_{n=0}^{U_l}\!e^{-\tilde{\Lambda}_l sT}\frac{(\tilde{\Lambda}_l sT)^n}{n!}\bv_l(n), l\!=\!0, 1, ...
      \label{eq: system_state_mp_u_t}
\end{equation}
where $\bv_l(n)=\bv_l(n-1)\tilde{\bP}_l$ and
$\bv_l(0)=\hat{\tilde{\bpi}}(lsT)$. The initial state is
$\hat{\tilde{\bpi}}(0)=\bpi(0)$.  The error is bounded as follows.
\begin{equation}
\hat{\tilde{\epsilon}}_l \!\leq\! \hat{\tilde{\epsilon}}_{l-1} \!+\!
\left(\!1\!\!-\!\!\sum\nolimits_{n=0}^{U_l}\!\!e^{-\tilde{\Lambda}_l sT}\frac{(\tilde{\Lambda}_l sT)^n}{n!}\!\right),
\ l\!=\!0,1,...
\label{eq: error_bound_mp_u_t}
\end{equation}
where $\hat{\tilde{\epsilon}}_0=||\hat{\tilde{\bpi}}(0) - \bpi(0)||_1=0$.

\label{theo: matrix_perturbation}
\end{theorem}

\noindent {\bf Proof:} Please refer to Appendix. \done

\subsection{Algorithm for Computing System State}

In the last subsection, we present the mathematical foundation on computing
the system state of SSD RAID arrays and the corresponding error bounds. We
now present the algorithm to compute $\hat{\tilde{\bpi}}(t)$ according to
Theorem~\ref{theo: matrix_perturbation}.  In particular, we aim to compute
the system state at the time when the $k^{th}$ erasure operation has just
occurred, i.e., $\hat{\tilde{\bpi}}(kT)$. Without loss of generality, we
assume that $k$ is an integer multiple of the step size $s$.  Moreover, we
denote the maximum acceptable error by $\epsilon$.
\begin{algorithm}
\caption{Algorithm for Computing System State $\hat{\tilde{\bpi}}(kT)$}
\label{alg: alg1}
\begin{small}
\begin{algorithmic}[1]
    \REQUIRE Step size $s$, maximum error $\epsilon$ and initial state
	$\hat{\tilde{\bpi}}(0) = \bpi(0)$
    \ENSURE System state at time $kT$: $\hat{\tilde{\bpi}}(kT)$
    \FOR {$l=0 \to \frac{k}{s}-1$}
    \STATE Let $\tilde{\bQ_l}=\frac{\sum_{m=ls}^{(l+1)s-1}\bQ_m}{s}$;
    \STATE Choose $\tilde{\Lambda}_l \geq
            \max_{ls\leq m < (l+1)s}\max_{0\leq i\leq S+1}|-q_{i,i}(m)|$;
    \STATE Let $\tilde{\bP}_l=\bI + \frac{\tilde{\bQ_l}}{\tilde{\Lambda}_l}$;
    \STATE Initialize: $\hat{\tilde{\epsilon}}_l\leftarrow 0$; $n \leftarrow 0$;
            $\hat{\tilde{\bpi}}((l+1)sT) \leftarrow \boldsymbol{0}$; $\bv_l(0)
           \leftarrow \hat{\tilde{\bpi}}(lsT)$;
    \WHILE {$1-\hat{\tilde{\epsilon}}_l >  \frac{s\epsilon}{k}$} \label{line: condition}
        \STATE $\hat{\tilde{\epsilon}}_l \leftarrow  \hat{\tilde{\epsilon}}_l + e^{-\tilde{\Lambda}_lsT}\frac{(\tilde{\Lambda}_l
                sT)^n}{n!}$;
        \STATE $\hat{\tilde{\bpi}}((l+1)sT) \leftarrow \hat{\tilde{\bpi}}((l+1)sT)  + e^{-\tilde{\Lambda}_l sT}\frac{(\tilde{\Lambda}_l sT)^n}{n!}\bv_l(n)$;
        \STATE $n \leftarrow n+1$;
        \STATE $\bv_l(n)\leftarrow \bv_l(n-1)\tilde{\bP}_l$; \label{line: mv_mul}
    \ENDWHILE
    \ENDFOR
\end{algorithmic}
\end{small}
\end{algorithm}

Algorithm~\ref{alg: alg1} describes the pseudo-code of the algorithm.
Lines~2 to 11 are to derive the system state in one interval with $s$ time
periods based on the flow in Section~\ref{subsec: analyze_nonhomogeneous}.
In particular, Line~2 constructs the generator matrix of our defined CTMC
$\{\tilde{X}(t)\}$. Lines~3 to 5 initialize the necessary parameters.
Lines~6 to 11 implement Equation~(\ref{eq: system_state_mp_u_t}), while the
truncation point is determined based on Equation~(\ref{eq:
error_bound_mp_u_t}) and the given maximum error. Note that the condition in
Line~\ref{line: condition} indicates that the maximum allowable error in one
interval is $\frac{s\epsilon}{k}$, as there are $\frac{k}{s}$ intervals and
the aggregate maximum allowable error is $\epsilon$.  After computing the
system state at time $kT$ using Algorithm~\ref{alg: alg1}, we can easily
compute the RAID reliability  based on the definition in Equation~(\ref{eq:
reliability_def}).

Our implementation of Algorithm~\ref{alg: alg1} uses the following inputs.
We fix $s=BM/20$, meaning that for each SSD, we consider at least 20 time
points before it reaches its lifetime of $BM$ erasures.  The error bound is
fixed at $\epsilon=10^{-3}$. We also set $\pi_0(0) = 1$ and $\pi_j(0)=0$ for
$0<j\le S+1$ to indicate that the array has no erroneous chunk initially.

Note that the dimension of the matrix $\tilde{\bP}_l$ is $(S+2) \times
(S+2)$ ($S$ is the number of stripes), which could be very large for large
SSDs. To further speed up our computation, we develop another optimization
technique by truncating the states with large state numbers from the CTMC so
as to reduce the dimension of $\tilde{\bP}_l$.  Intuitively, if an array
contains many stripes with exactly one erroneous chunk, it is more likely
that a new erroneous chunk appears in one of such stripes (and hence data
loss occurs) rather than in a stripe without any erroneous chunk. That is,
the transition rate $q_{i,i+1}$ becomes very small when $i$ is large.  We
can thus remove such states with large state numbers without losing
accuracy. We present the details of the optimization technique in the next
subsection.

\subsection{Reducing Computational Cost of Algorithm \ref{alg: alg1}}

\begin{figure}
  \centering
  \includegraphics[width=7cm]{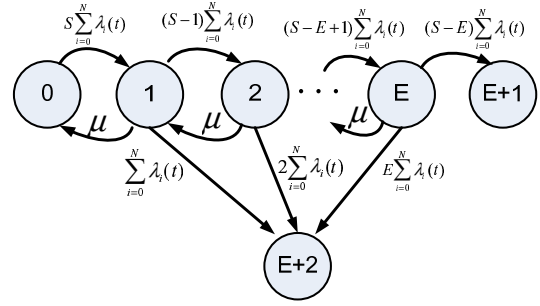}\\
  \caption{State transition after  truncation.}\label{fig: states_truncation}
\end{figure}

Note that when  state number $i$ increases, the transition rate
$q_{i,i+1}(k)$ decreases while the transition rate $q_{i,S+1}(k)$ increases.
This indicates that the higher the current state number is, the harder it is
to transit to states with larger state number, while it is easier to transit
to the state of data loss, or state $S+1$. The physical meaning is that the
system will not contain too many stripes with exactly one erroneous chunk as
either the erroneous chunk will be recovered, or another error may appear in
the same stripe so that data loss happens. Therefore, to reduce the
computational cost when derive the system state, we can truncate the states
with large state number so as to reduce the state space of the Markov chain.
Specifically, we truncate the states with state number bigger than $E$, and
let  $E\!+\!1$  represents the case when more than $E$ stripes contain
exactly one erroneous chunk. Moreover, we take  state $E+1$ as an absorbing
state. Furthermore, we denote the state of data loss by $E+2$. Now, the
state transition can be illustrated in Figure \ref{fig: states_truncation}.

To compute the system state after states truncation, we denote the new CTMC
by $\{\bar{X}(t), t\geq 0\}$,   the new  generator matrix during period
$(kT, (k+1)T)$ by $\bar{\bQ}_k$, and  the system state at time $(k+1)T$  by
$\bar{\bpi}((k+1)T)$. We use notations with a bar to represent the case when
system states of the CTMC are truncated if the context is clear. Similar to
Equation~(\ref{eq: system_state_mp}), given the
 initial state $\bar{\bpi}(kT)$, the system state at time
$(k+1)T$ for the CTMC  $\{\bar{X}(t), t\geq 0\}$ can be derived as follows.
\begin{equation}
  \bar{\bpi}((k+1)T)=\bar{\bpi}(kT)\sum_{n=0}^{\infty}\frac{(\bar{\bQ}_kT)^n}{n!}.
\label{eq: system_state_statetruncation}
\end{equation}
If we denote the error  caused by truncating the states at time $kT$ by
$\bar{\epsilon}_k$,  then $\bar{\epsilon}_k$ can be formally defined as
follows.
\begin{equation*}
 \bar{\epsilon}_k =\max_{0\leq i\leq E}
|\bar{\pi}_i(kT)-\pi_i(kT)|,
\end{equation*}
where $\bar{\pi}_i(kT)$ represents the probability of system being at state
$i$ at time $kT$ for the CTMC $\{\bar{X}(t), t \geq 0\}$, i.e., the Markov
chain after states truncation, and $\pi_i(kT)$ represents the probability of
the system being at state $i$ at time $kT$ for the original CTMC $\{X(t), t
\geq 0\}$. Clearly, $\bar{\epsilon}_0=0$ as the two Markov chains have the
same initial states, i.e., $\bar{\pi}_i(0)=\pi_i(0)$. The bound of the error
caused by states truncation is
\begin{equation}
\bar{\epsilon}_k \leq \bar{\pi}_{E+1}(kT).
\label{eq: error_bound_statetruncation}
\end{equation}

Again, we can also follow the steps in Section \ref{subsec:
analyze_nonhomogeneous}, i.e., use Algorithm \ref{alg: alg1}, to compute the
system state for the Markov chain after states truncation $\{\bar{X}(t), t\!
\geq\! 0\}$.

\section{{\bf Model Validation}}
\label{sec: validation}

In this section, we validate via trace-driven simulation the accuracy of our
CTMC model on quantifying the RAID reliability $R(t)$. We use the
Microsoft's SSD simulator \cite{Agrawal08} based on DiskSim \cite{disksim}.
Since each SSD contains multiple chips that can be configured to be
independent of each other and handle I/O requests in parallel, we consider
RAID at the chip level (as opposed to device level) in our DiskSim
simulation.  Specifically, we configure each chip to have its own data bus
and control bus and treat it as one drive, and also treat the SSD controller
as the RAID controller where parity-based RAID is built.

To simulate error arrivals, we generate error events based on Poisson
arrivals given the current system age $k$ of the array.  As the array ages,
we update the error arrival rates accordingly by varying the variable
$k_i(t)$ in Equation~(\ref{eq: error_rate}).  We also generate recovery
events whose recovery times follow an exponential distribution with a fixed
rate $\mu = 1$.  Both error and recovery events are fed into the SSD
simulator as special types of I/O requests.  We consider three cases: {\em
error dominant}, {\em comparable}, and {\em recovery dominant}, in which the
error rate is larger than, comparable to, and smaller than the recovery
rate, respectively.

Our validation measures the reliability of the traditional RAID and
Diff-RAID with different parity distributions.  Recall that Diff-RAID
redistributes the parities after each drive replacement, while traditional
RAID does not.  We consider ($N+1$) chips where $N=3,5,7$.  For traditional
RAID, we choose RAID-5, in which parity chunks are evenly placed across the
chips; for Diff-RAID, 10\% of parity chunks placed in each of the $N$ chips
and the remaining parity chunks are placed in the last flash chip.

We generate synthetic uniform workload in which the write requests access
the addresses of the entire address space with equal probability.  The
workload lasts until all drives are worn out and replaced at least once.  We
run the DiskSim simulation 1000 times, and in each run we record the age
when data loss happens. Finally, we derive the probability of data loss and
the reliability based on our definitions.  To speed up our DiskSim
simulation, we consider a small-scale RAID array, in which each chip
contains 80 blocks with 64 pages each, and the chunk size is set to be the
same as the page size 4KB. We also set a low erasure limit at $M=100$ cycles
for each block.

\begin{figure*}[!t]
  \centering
  \subfloat[Error dominant case (3+1 RAID)]{
  \includegraphics[width=0.3\textwidth]{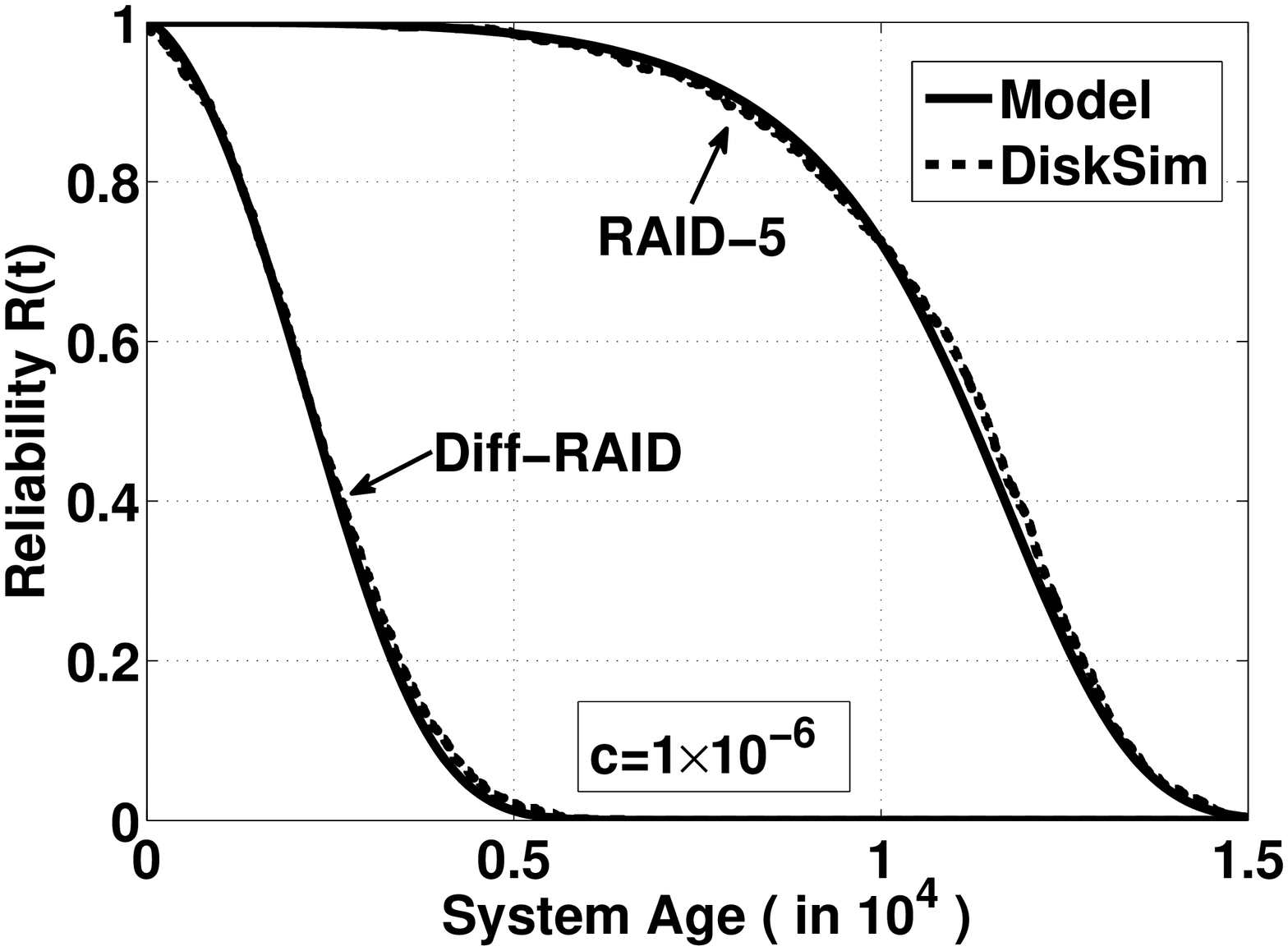}}
  \subfloat[Comparable case (3+1 RAID)]{
  \includegraphics[width=0.3\textwidth]{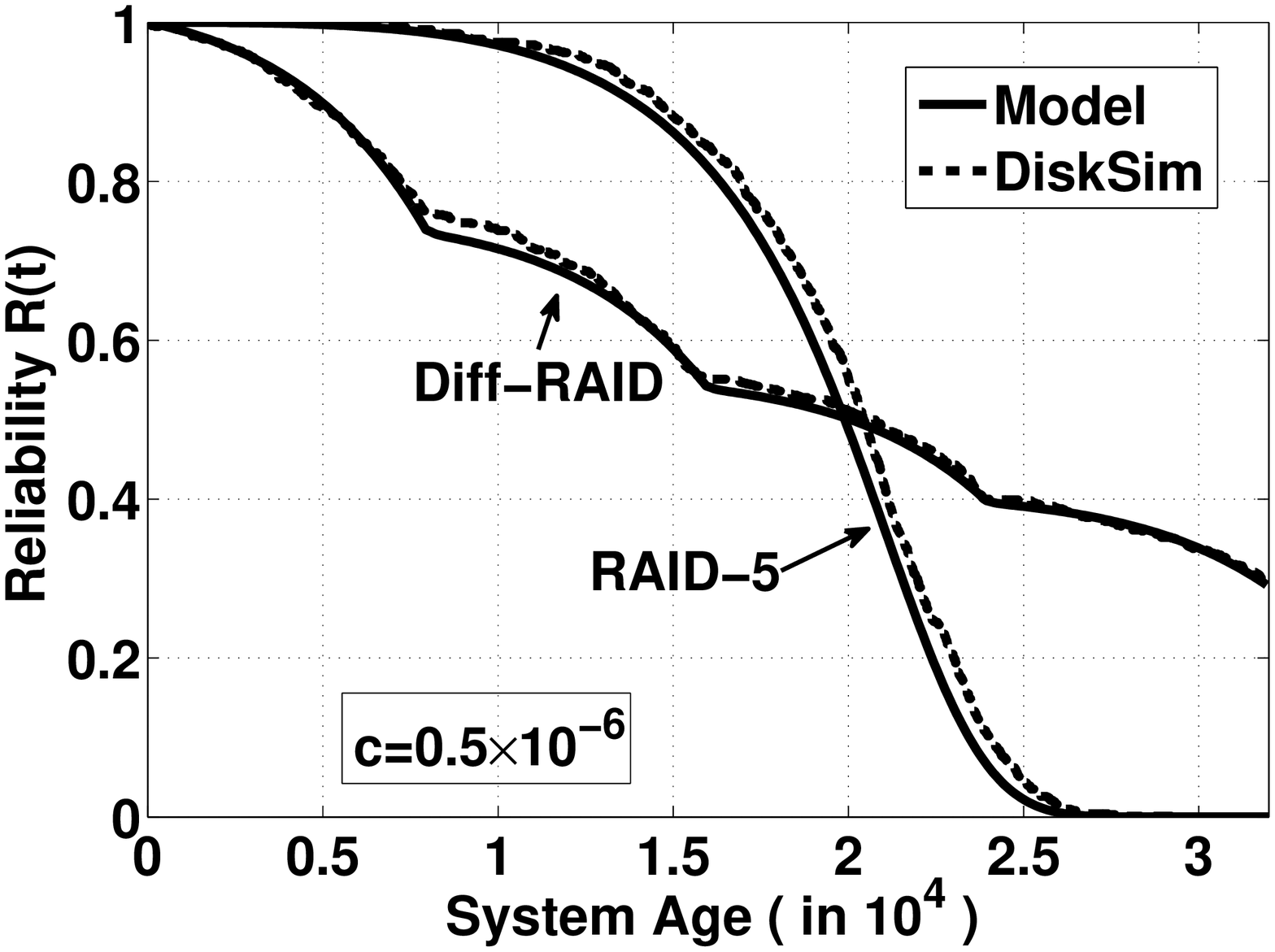}}
  \subfloat[Recovery dominant case (3+1 RAID)]{
  \includegraphics[width=0.3\textwidth]{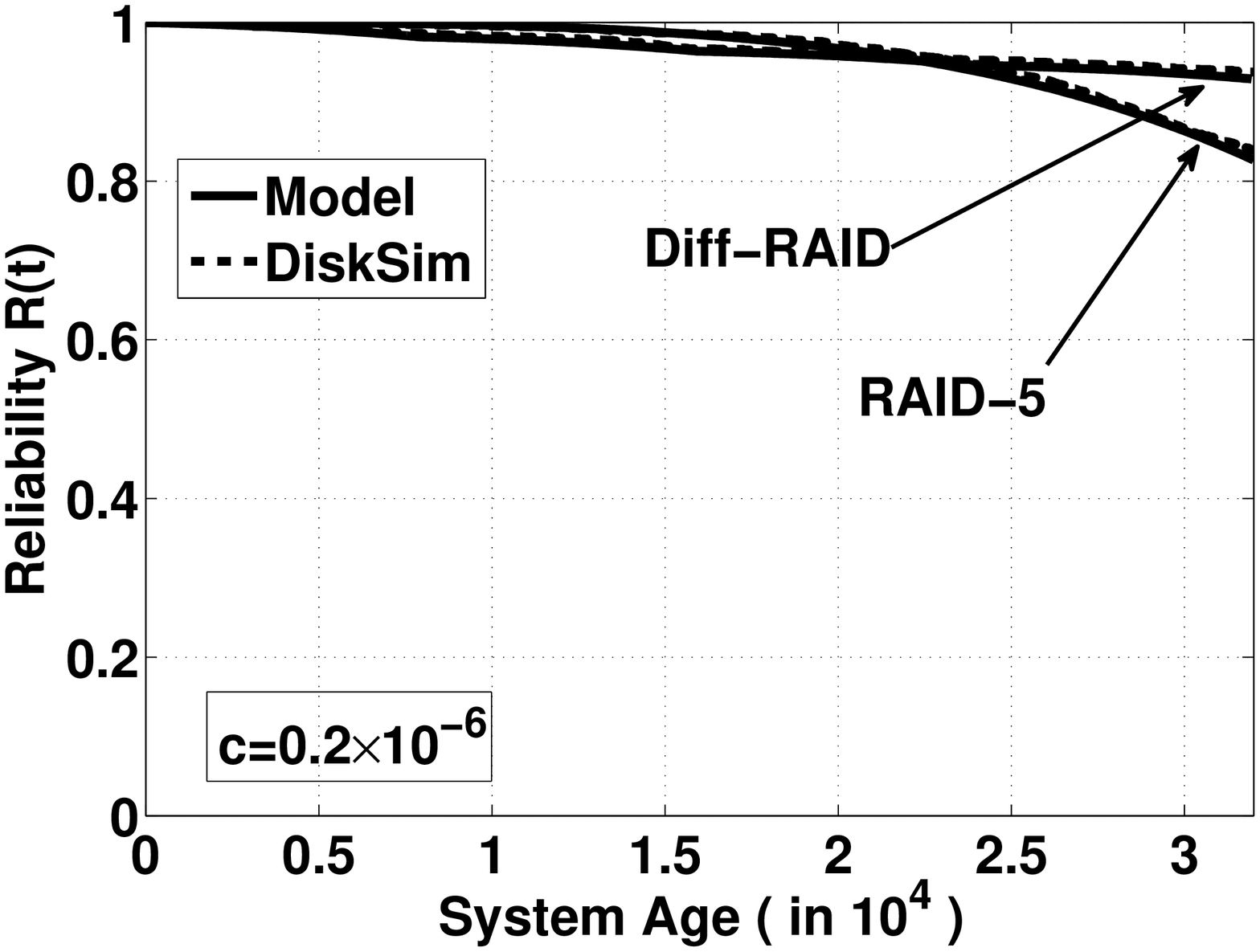}}\\
  \subfloat[Error dominant case (5+1 RAID)]{
  \includegraphics[width=0.3\textwidth]{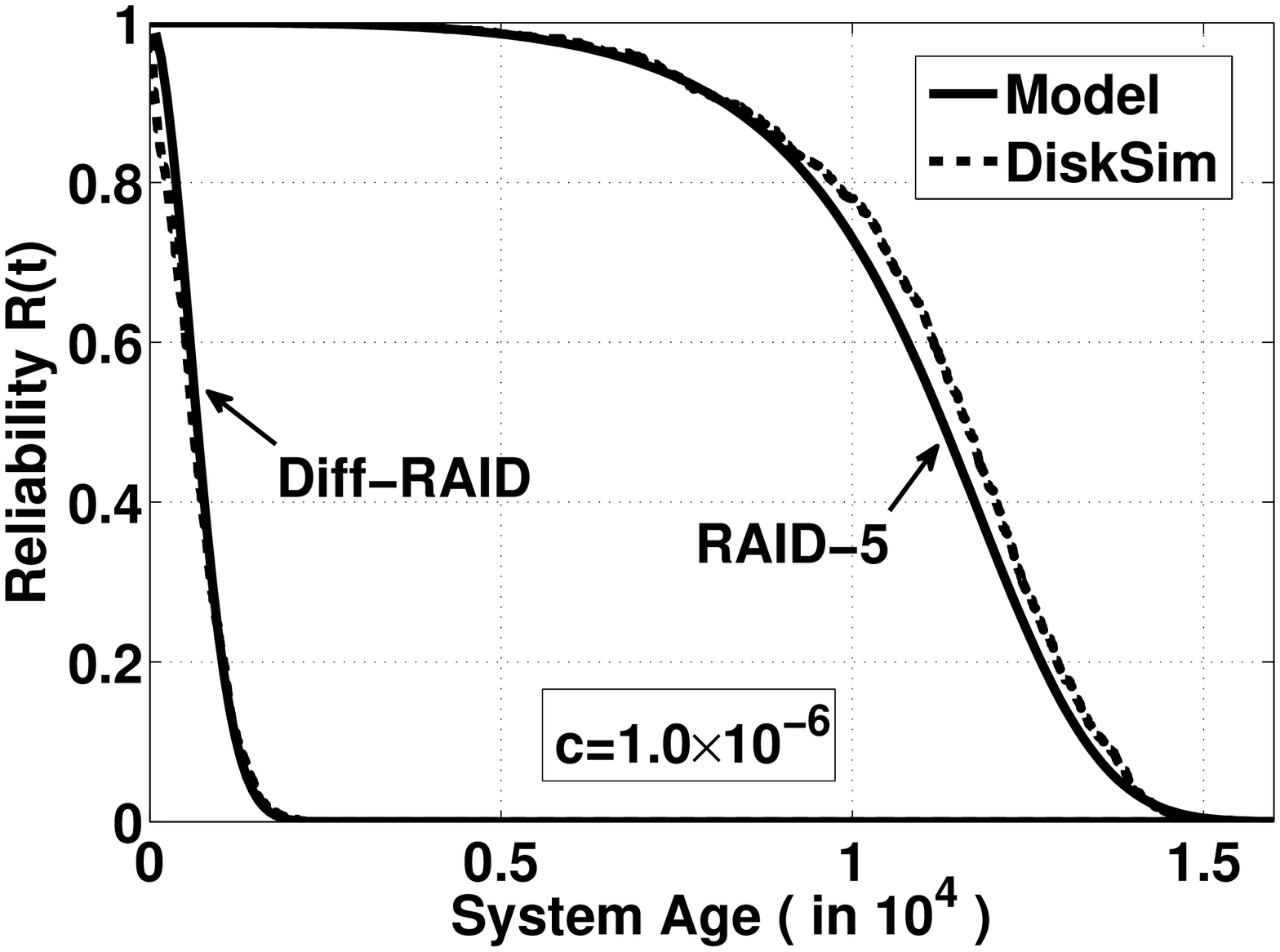}}
  \subfloat[Comparable case (5+1 RAID)]{
  \includegraphics[width=0.3\textwidth]{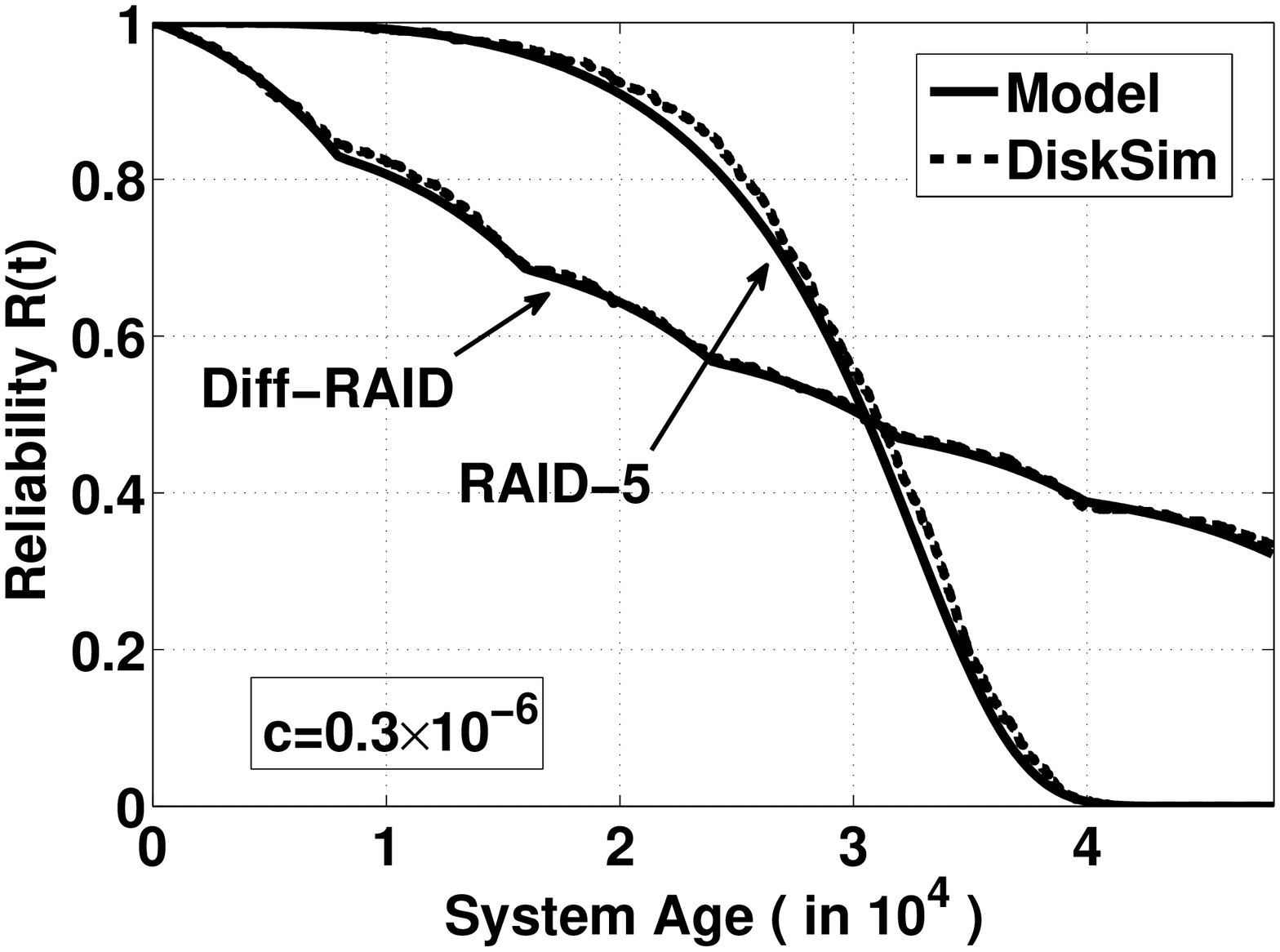}}
  \subfloat[Recovery dominant case (5+1 RAID)]{
  \includegraphics[width=0.3\textwidth]{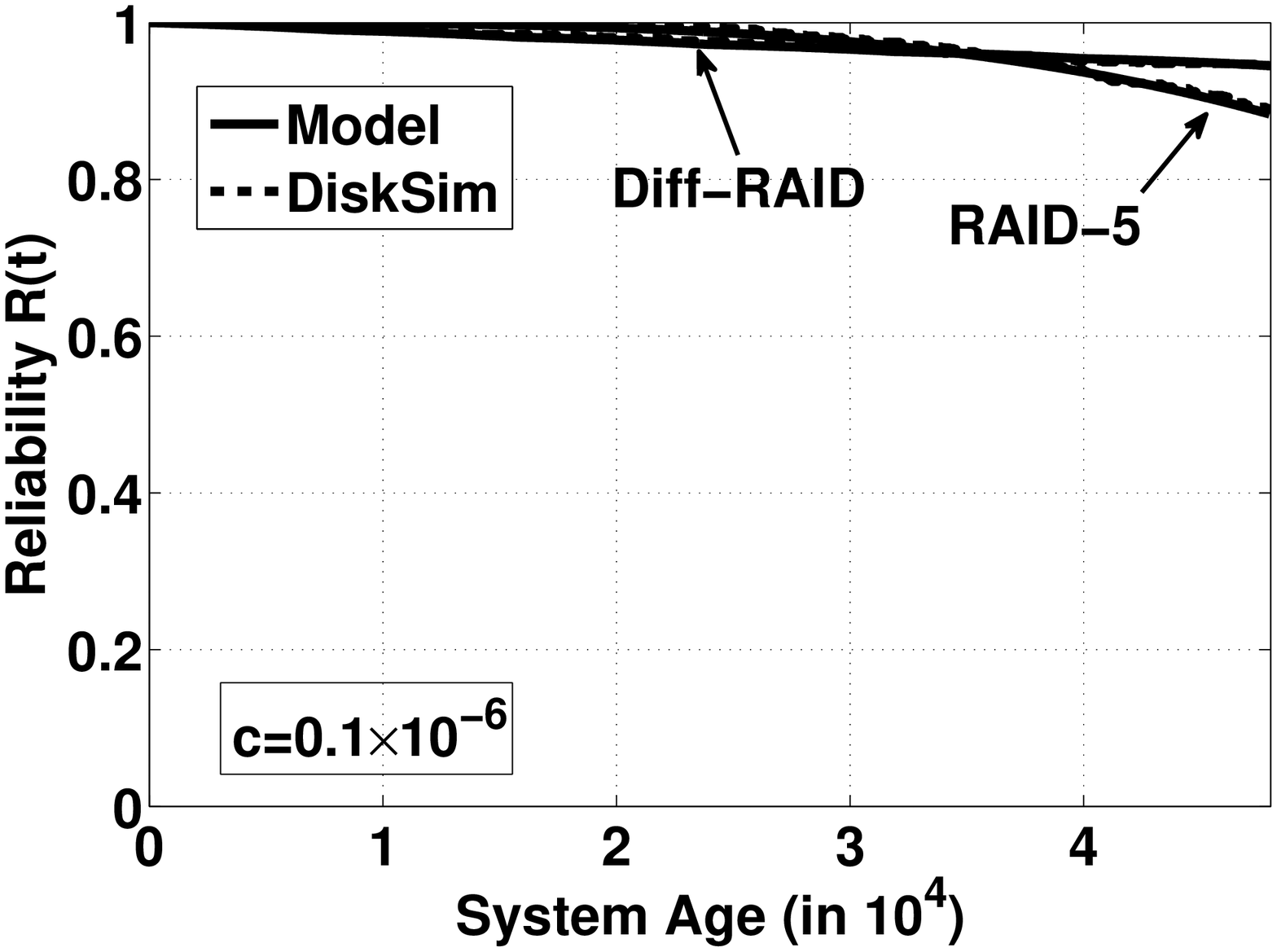}}\\
  \subfloat[Error dominant case (7+1 RAID)]{
  \includegraphics[width=0.3\textwidth]{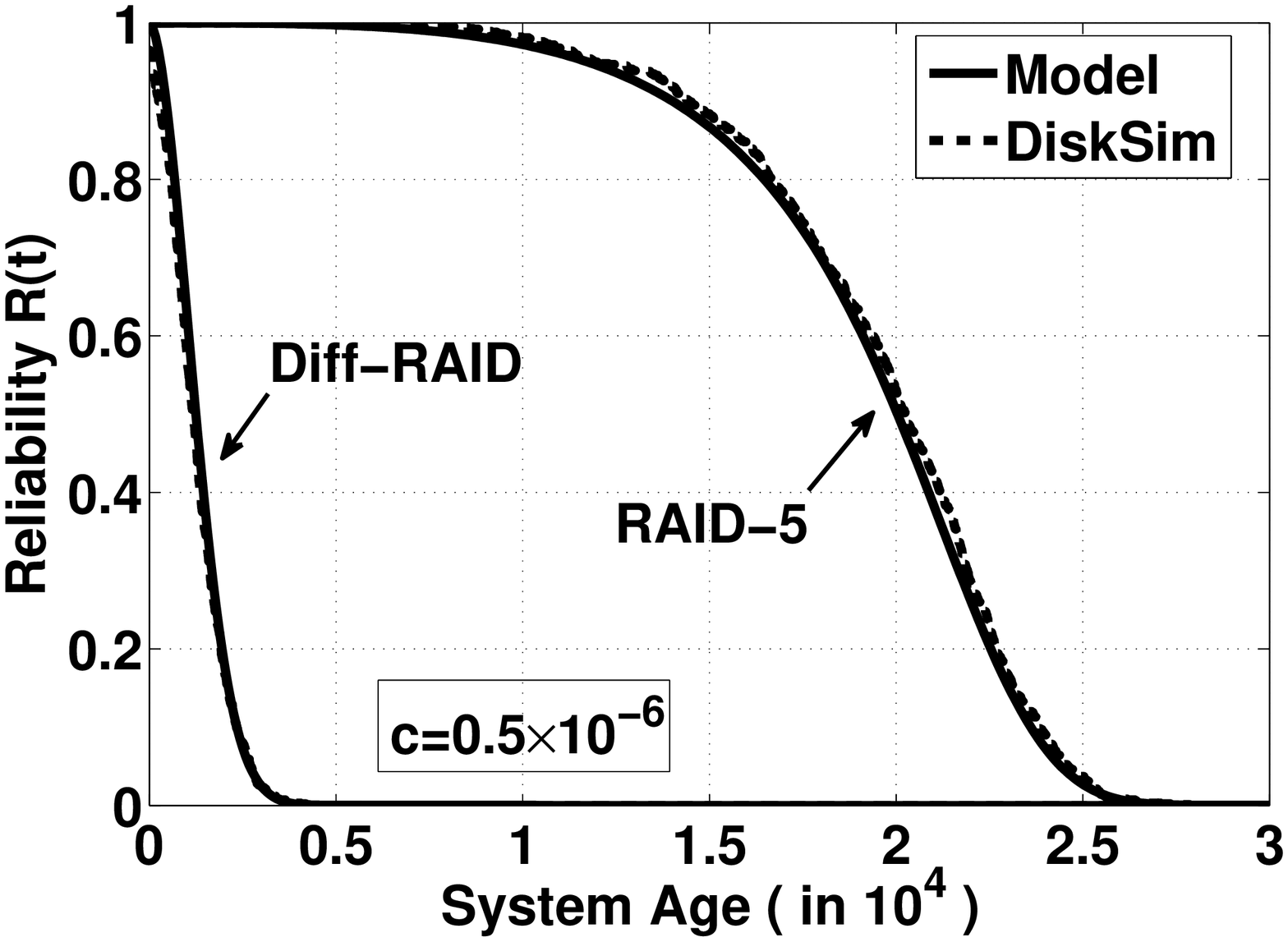}}
  \subfloat[Comparable case (7+1 RAID)]{
  \includegraphics[width=0.3\textwidth]{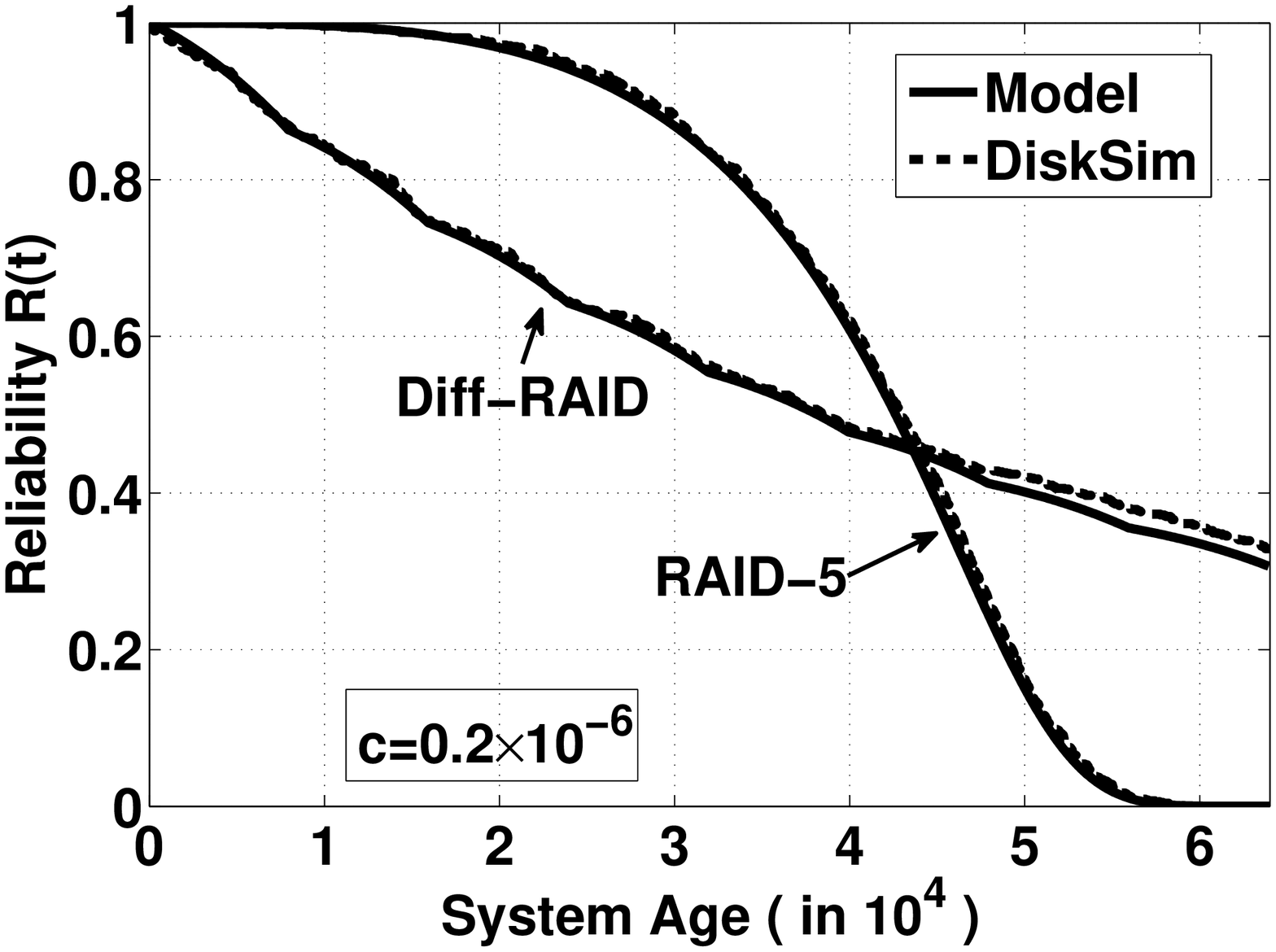}}
  \subfloat[Recovery dominant case (7+1 RAID)]{
  \includegraphics[width=0.3\textwidth]{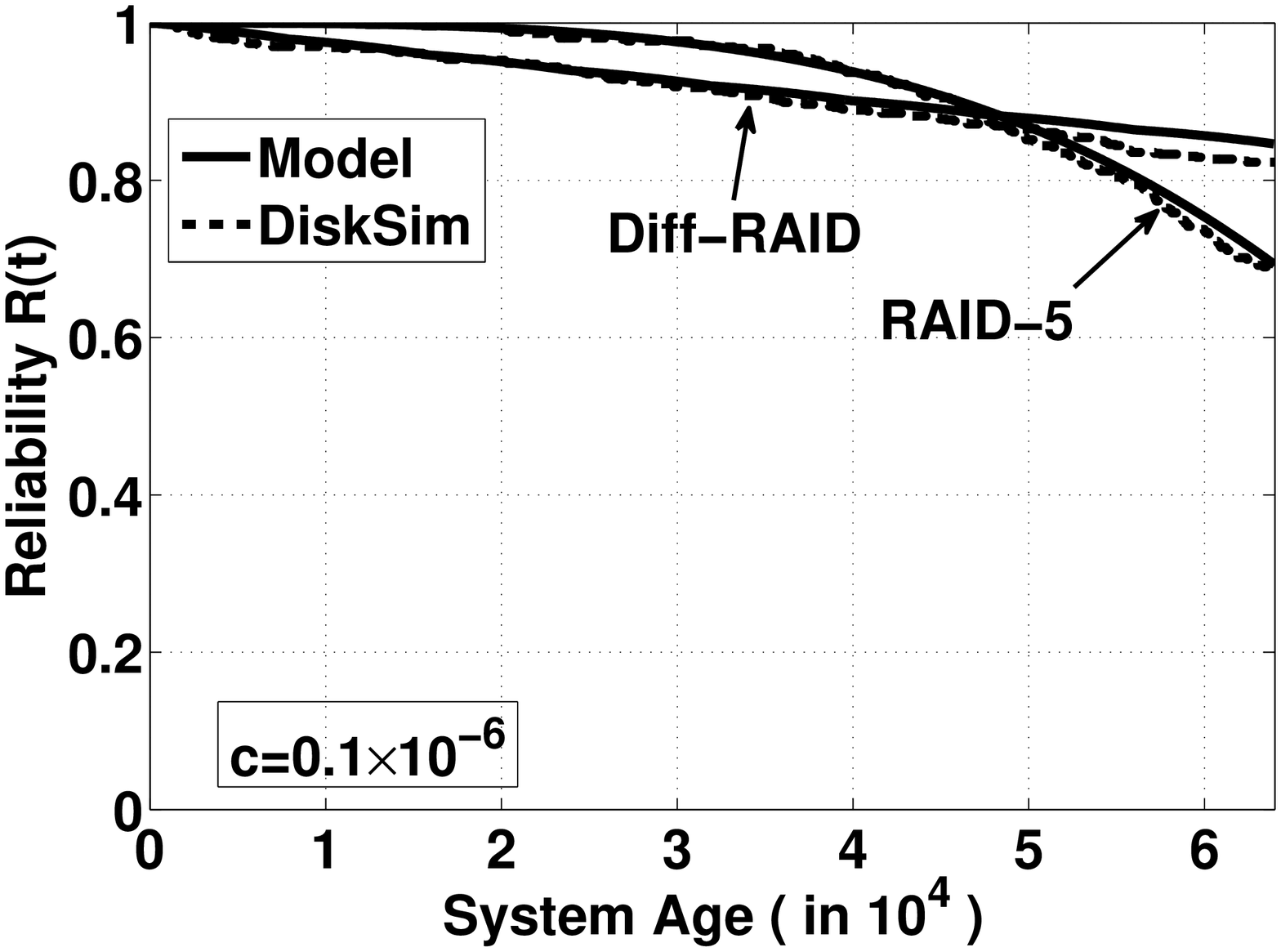}}
  \caption{Model validation with respect to different values of $N$ and
  different error rates.}
  \label{fig: validation}
\end{figure*}

Figure~\ref{fig: validation} shows the reliability $R(t)$ versus the system
age $k$ obtained from both the model and DiskSim results.  We observe that
our model accurately quantifies the reliability for all cases.  Also,
Diff-RAID shows its benefit only in the comparable case.  In the error
dominant case, traditional RAID always shows higher reliability than
Diff-RAID; in the recovery dominant case, there is no significant difference
between traditional RAID and Diff-RAID.  We will further discuss these
findings in Section~\ref{sec: simulation}.

\section{{\bf Numerical Analysis}}
\label{sec: simulation}

In this section, we conduct numerical analysis on the reliability dynamics
of a large-scale SSD RAID array with respect to different parity placement
strategies. To this end, we summarize the lessons learned from our analysis.

\subsection{Choices of Default Model Parameters}\label{subsec: parameter}

We first describe the default model parameters used in our analysis, and
provide justifications for our choices.

We consider an SSD RAID array composed of $N+1$ SSDs, each being modeled by
the same set of parameters.  By default, we set $N = 9$. Each block of an
SSD has 64 pages of size 4KB each. We consider 32GB SSDs with $B = 131,072$
blocks. We configure the chunk size equal to the block size, i.e., there are
$S = B = 131,072$ chunks\footnote{In practice, SSDs are over-provisioned
\cite{Agrawal08}, so the actual number of blocks (or chunks) that can be
used for storage (i.e., $S$) should be smaller. However, the key
observations of our results here still hold.}.  We also have each block
sustain $M = $10K erase cycles.

We now describe how we configure the error arrival rate, i.e.,
$\lambda_i=2ck_i$, by setting the constant $c$.  We employ 4-bit ECC
protection per 512 bytes of data, the industry standard for today's MLC
flash. Based on the uncorrectable bit error rates (UBERs) calculated in
\cite{Balakrishnan10}, we choose the UBER in the range $[10^{-16},10^{-18}]$
when an SSD reaches its rated lifetime (i.e., the erasure limit $M$ is
reached).  Since we set the chunk size to be equal to the block size, the
probability that a chunk contains at least one bit error is roughly in the
range of $[2\times10^{-10}, 2\times10^{-12}]$.  Based on the analysis on
real enterprise workload traces \cite{Narayanan09}, an RAID array can have
several hundred gigabytes of data being accessed per day.  If the write
request rate is set as 1TB per day (i.e., 50 blocks per second), then the
error arrival rate per chunk at its rated lifetime (i.e., $\lambda_i = 2cM$)
is approximately in the range $[10^{-8}, 10^{-10}]$. The corresponding
parameter $c$ is in the range $[0.5\times 10^{-12}, 0.5 \times 10^{-14}]$.

For the error recovery rate $\mu$, we note that the aggregate error arrival
rate when all $N+1$ drives are going to die out is $2cMS(N+1)$.  If $N=9$,
then the aggregate error arrival rate is roughly in the range $[10^{-2},
10^{-4}]$.  We fix $\mu = 10^{-3}$.

We compare different cases when the error arrivals are more dominant than
error recoveries, and vice versa.  We consider three cases of error
patterns: $c = 1.1\times 10^{-13}$, $c = 0.4\times 10^{-13}$, and $c =
0.1\times 10^{-13}$, which correspond to the {\em error dominant},  {\em
comparable}, and {\em recovery dominant} cases, respectively.  Specifically,
when $c = 0.4\times 10^{-13}$, the aggregate error arrival rate of the array
when all SSDs reach their rated lifetime is around $2cMS(N+1) \approx
10^{-3}$ (where $N=9$, $M=$10K, and $S=131,072$).

We now configure $T$, the time interval between two neighboring erase
operations.  Suppose that there are 1TB of writes per day as described
above. The inter-arrival time of write requests is around $3\times 10^{-4}$
seconds for 4KB page size.  Thus, the average time between two erase
operations is $1.9\times10^{-2}$ seconds as an erase is triggered after
writing 64 pages. In practice, each erase causes additional writes (i.e.,
write amplification \cite{Hu09}) as it moves data across blocks, so $T$
should be smaller.  Here, we fix $T=10^{-2}$ seconds.

We compare the reliability dynamics of RAID-5 and different variants of
Diff-RAID.  For RAID-5, each drive holds a fraction $\frac{1}{N+1}$ of
parity chunks; for Diff-RAID, we choose the parity distribution (i.e.,
$p_i$'s for $0\le i\le N$) based on a truncated normal distribution.
Specifically, we consider a normal distribution $\mathcal{N}(N+1, \sigma^2)$
with mean $N+1$, and standard deviation $\sigma$, and let $f$ be the
corresponding probability density function.  We then choose $p_i$'s as
follows:
\begin{equation}
  p_i=\frac{\int_{i}^{i+1}f(x)dx}{\int_{0}^{N+1}f(x)dx}, \quad 0 \leq i \leq N.
  \label{eq: parity_dist_sim}
\end{equation}
We can choose different distributions of $p_i$ by tuning the parameter
$\sigma$.  Intuitively, the larger $\sigma$ is, the more evenly $p_i$'s are
distributed.  We consider three cases: $\sigma=1$, $\sigma=2$, and
$\sigma=5$.  Suppose that $N = 9$. Then for $\sigma=1$, SSD~$N$ and
SSD~$N-1$ hold 68\% and 27\% of parity chunks, respectively;  for
$\sigma=2$, SSD~$N$, SSD~$N-1$, and SSD~$N-2$ hold 38\%, 30\%, and 18\% of
parity chunks, respectively; for $\sigma=5$, the proportions of parity
chunks range from 2.8\% (in SSD~0) to 16.6\% (in SSD~$N$).  After choosing
$p_i$'s, the age of each block of SSD~$i$ (i.e., $k_i$) can be computed via
Equation~(\ref{eq: D-RAIDblock_age2}).

\subsection{Impact of Different Error Dynamics}
\label{subsec: reliability}

We now show the numerical results of RAID reliability based on the
parameters described earlier. We assume that drive replacement can be
completed immediately after the oldest SSD reaches its rated lifetime.  When
the oldest drive is replaced, all its chunks (including any erroneous
chunks) are copied to the new drive. Thus, the reliability (or the
probability of no data loss) remains the same. We consider three error
cases: error dominant,  comparable, and recovery dominant cases, as
described above.

\begin{figure*}[!t]
  \centering
  \subfloat[Error dominant case ($c=1.1\times 10^{-13}$) ]{
  \includegraphics[width=0.3\textwidth]{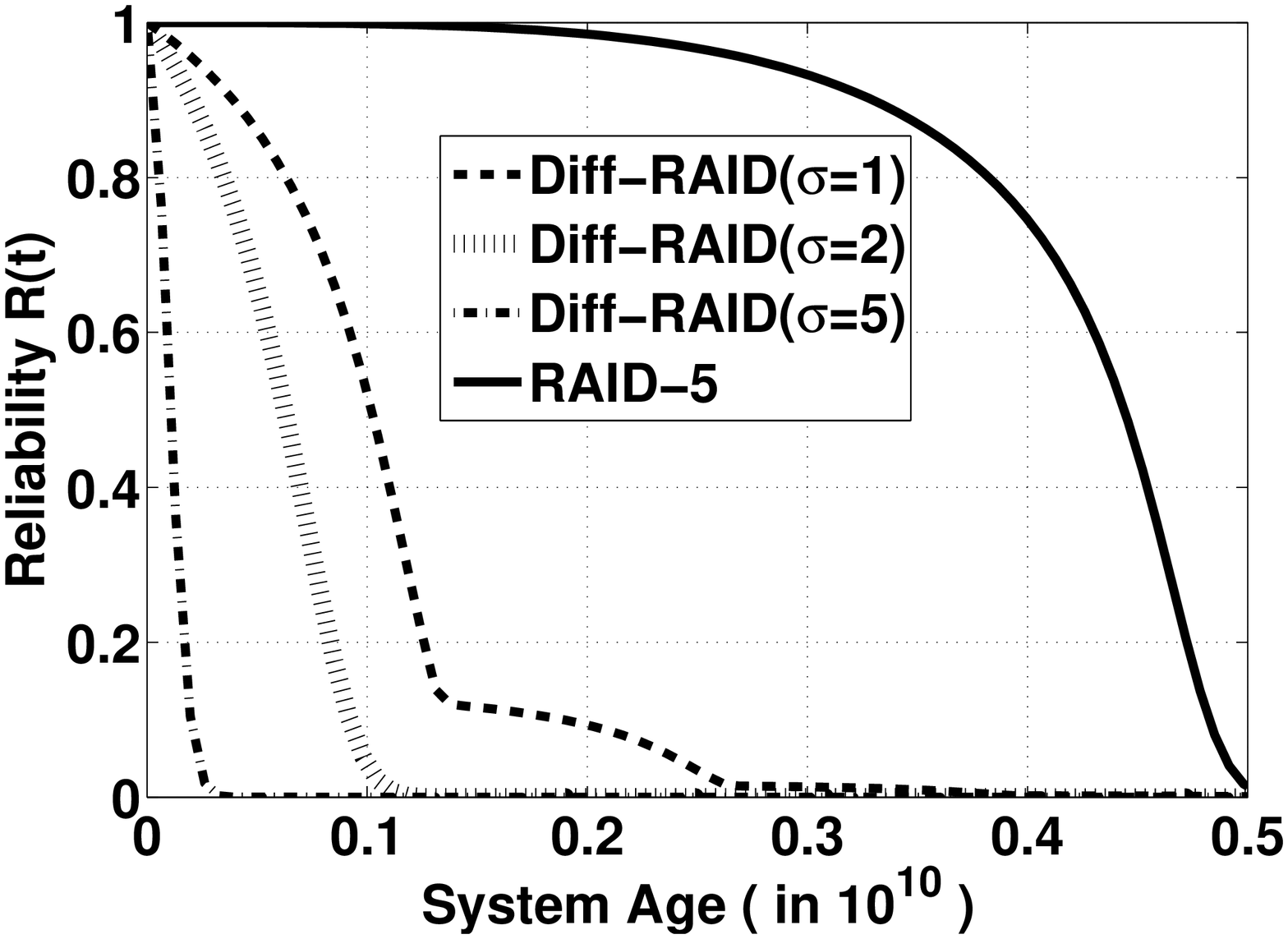}}
  \subfloat[Comparable case ($c=0.4\times 10^{-13}$)]{
  \includegraphics[width=0.3\textwidth]{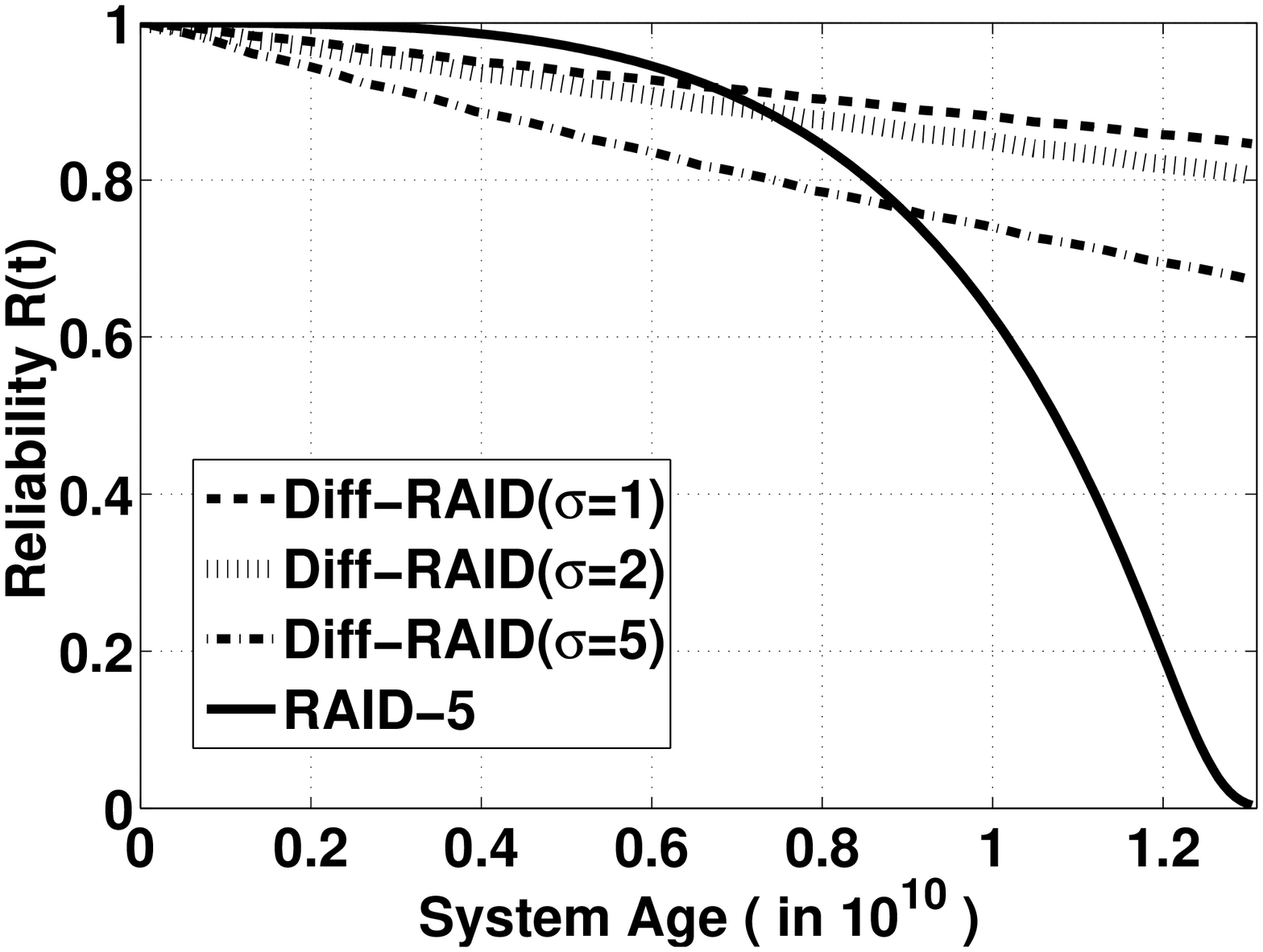}}
   \subfloat[Recovery dominant case ($c=0.1\times 10^{-13}$)]{
  \includegraphics[width=0.3\textwidth]{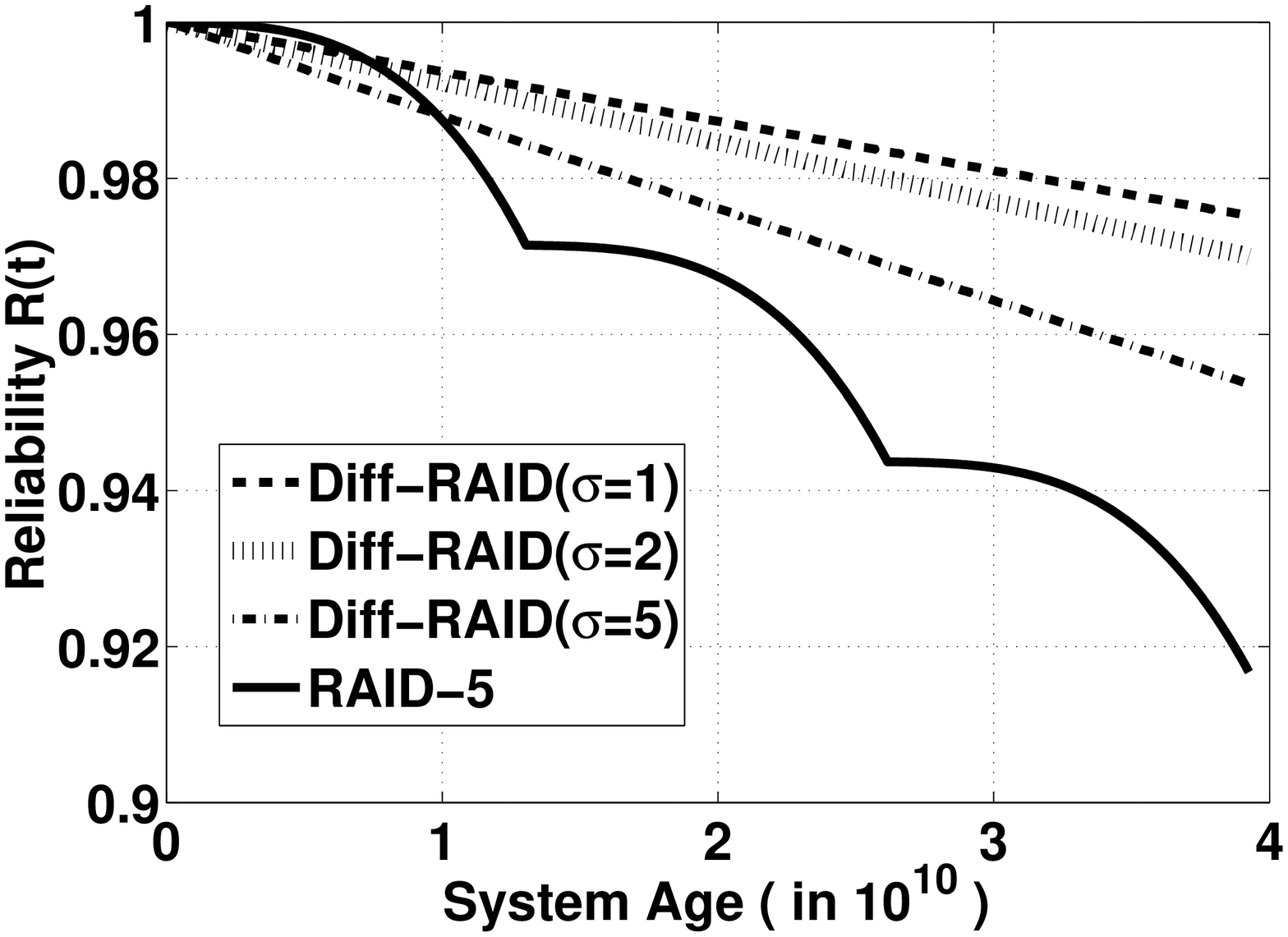}}
  \caption{Reliability dynamics of SSD arrays.}
  \label{fig: reliability_dynamics}
  \vspace{-15pt}
\end{figure*}

\noindent {\bf Case 1: Error dominant case.} Figure~\ref{fig:
reliability_dynamics}(a) first shows the numerical results for the error
dominant case.   Initially, RAID-5 achieves very good reliability as all
drives are brand-new. However, as SSDs wear down, the bit error rate
increases, and this makes the RAID reliability decrease very quickly. In
particular, the reliability drops to zero (i.e., data loss always happen)
when the array performs around $5\times10^{9}$ erasures. For Diff-RAID,  the
more evenly parity chunks are distributed, the lower RAID reliability is. In
the error dominant case, since error arrival rate is much bigger than the
recovery rate, the RAID reliability drops to zero very quickly no matter
what parity placement strategy is used.  We note that Diff-RAID is less
reliable than traditional RAID-5 in the error dominant case. The reason is
that for Diff-RAID, the initial ages of SSDs when constructing the RAID
array are non-zero, but instead follow the convergent age distribution
(i.e., based on $A_i$'s in Equation~(\ref{eq: age_dist_convergent})). When
error arrival rate is very large, the array suffers from low reliability
even if the array only performs small number of erasures. However, for
RAID-5, since it is always constructed by using brand-new SSDs, it starts
with a very high reliability.

\noindent {\bf Case 2: Comparable case.} Figure~\ref{fig:
reliability_dynamics}(b) shows the results for the comparable case. RAID-5
achieves very good reliability initially, but decreases dramatically as the
SSDs wear down.  Also, all drives wear down at the same rate, the
reliability of the array is about zero when all drives reach their erasure
limits, i.e., when the system age is around $1.3\times10^{10}$ erasures.
Diff-RAID shows different reliability dynamics.  Initially, Diff-RAID has
less reliability than RAID-5, but the drop rate of the reliability is much
slower than that of RAID-5 as SSDs wear down.  The reason is that Diff-RAID
has uneven parity placement, SSDs are worn out at different times and will
be replaced one by one.  When the worn-out SSD is replaced, other SSDs
perform fewer erase operations and have small error rates. This prevents the
whole array suffering from a very large error rate as in RAID-5.  Also, the
reliability is higher when the parity distribution is more skewed (i.e.,
smaller $\sigma$), as also observed in \cite{Balakrishnan10}.

\noindent {\bf Case 3: Recovery dominant case.} Figure~\ref{fig:
reliability_dynamics}(c) shows the results for the recovery dominant case.
RAID-5 shows high reliability in general.  Between two replacements (which
happens every $1.3\times10^{10}$ erasures), its data loss probability drops
by within 3\%.  Its reliability drops slowly right after each replacement,
and its drop rate increases as it is close to be worn out.  Diff-RAID shows
higher reliability than RAID-5 in general, but the difference is small
(e.g., less than 6\% between Diff-RAID for $\sigma=1$ and RAID-5).
Therefore, in the recovery dominant scenario, we may deploy RAID-5 instead
of Diff-RAID, as the latter introduces higher costs in parity redistribution
in each replacement and has smaller I/O throughput due to load imbalance of
parities.

\subsection{Impact of Different Array Configurations}

We further study via our model how different array configurations affect the
RAID reliability.  We focus on Diff-RAID and generate the parity
distribution $p_i$'s with $\sigma=1$.  Our goal is to validate the
robustness of our model on characterizing the reliability for different
array configurations.

\begin{figure*}[!t]
  \centering
  \subfloat[Impact of $N$ ]{
  \includegraphics[width=0.3\textwidth]{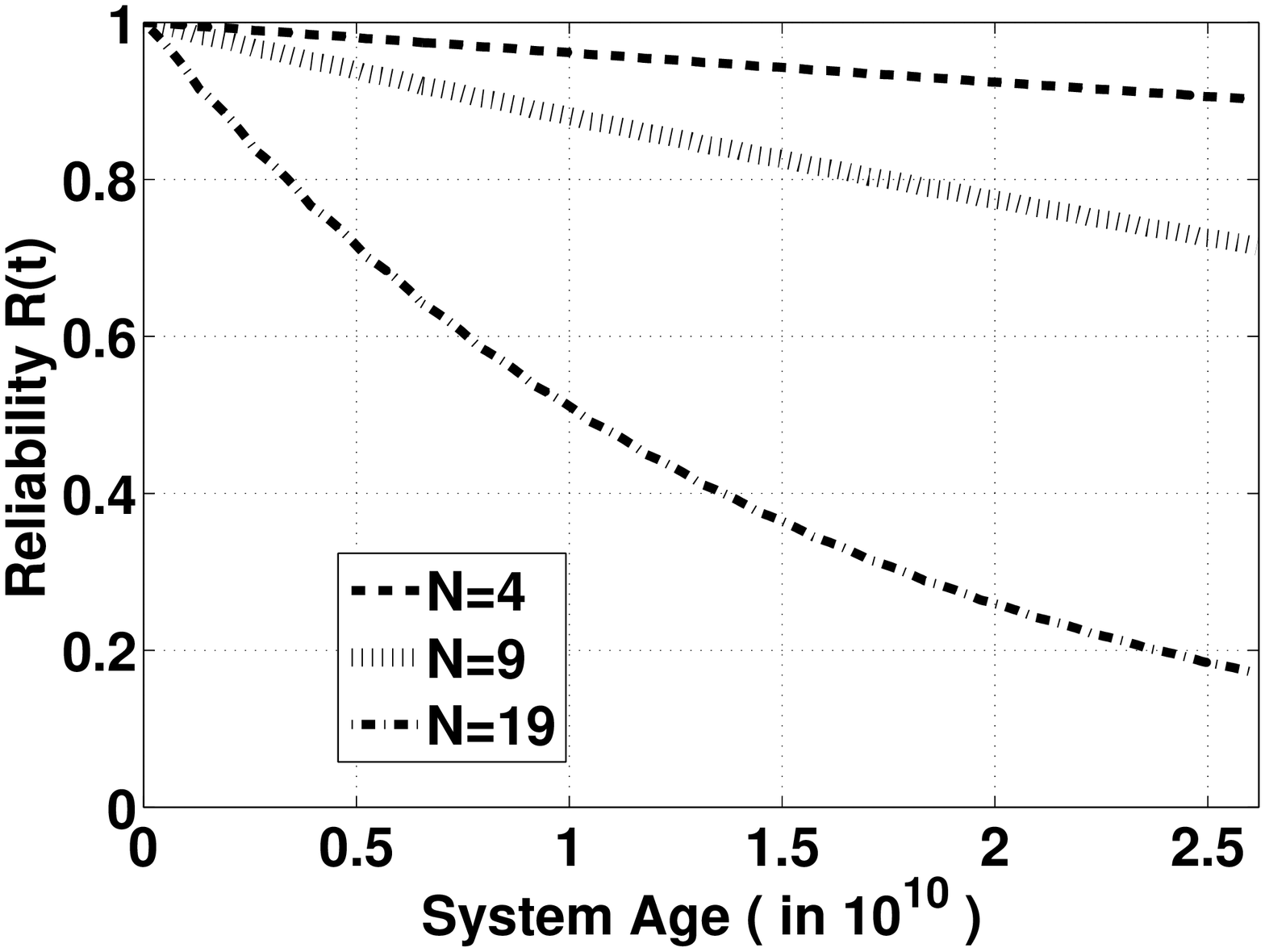}}
  \subfloat[Impact of ECC length]{
  \includegraphics[width=0.3\textwidth]{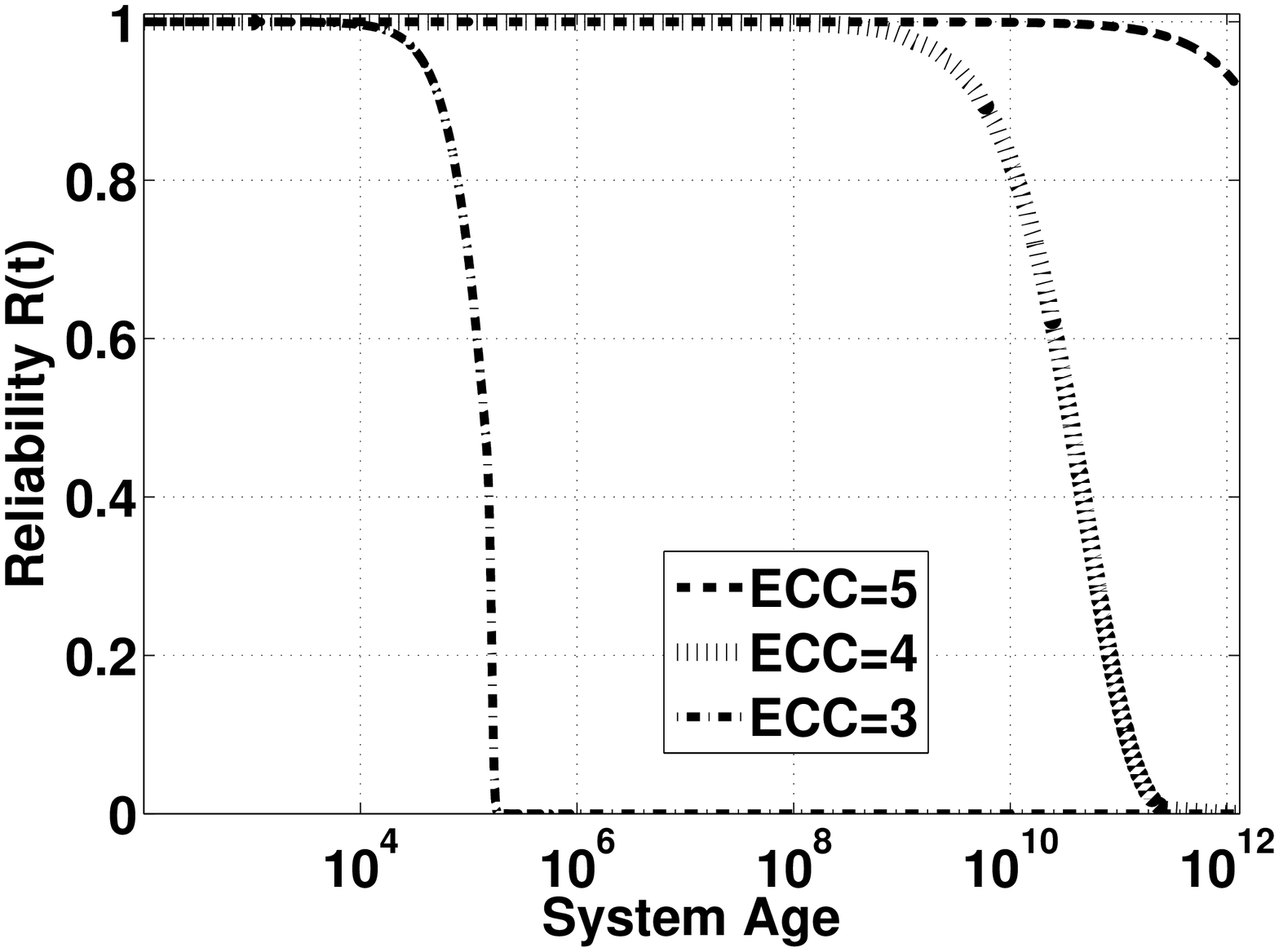}}
   \subfloat[Impact of $M$]{
  \includegraphics[width=0.3\textwidth]{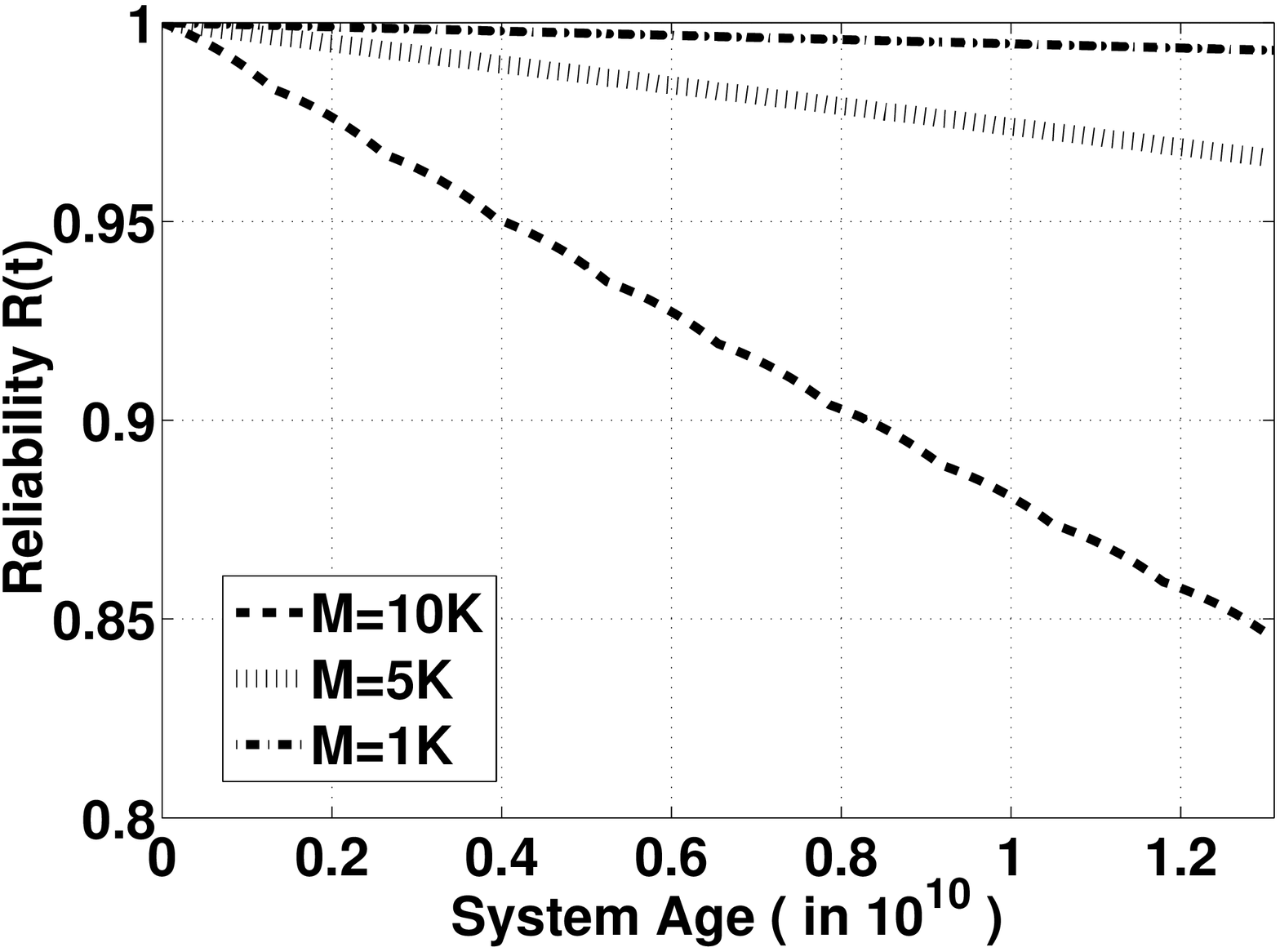}}
  \caption{Impact of different RAID configurations on the reliability.}
  \label{fig: array_conf}
\end{figure*}

\noindent {\bf Impact of $N$.} Figure~\ref{fig: array_conf}(a) shows the
impact of the RAID size $N$.  We fix other parameters as the same in the
comparable case, i.e., $\mu=10^{-3}$, $c=0.4\times10^{-13}$, and $M=10^4$.
The larger the system size, the lower the RAID reliability.  Intuitively,
the probability of having one more erroneous chunk in a stripe increases
with the stripe width (i.e., $N+1$).  Note that the reliability drop is
significant when $N$ increases.  For example, at $2.6\times10^{10}$
erasures, the reliability drops from 0.7 to 0.2 when $N$ increases from 9 to
19.

\noindent {\bf Impact of ECC.} Figure~\ref{fig: array_conf}(b) shows the
impact of different ECC lengths.  We fix $\mu=10^{-3}$, $M=10^4$, and $N=9$.
We also fix the raw bit error rate (RBER) as $1.3\times10^{-6}$
\cite{Balakrishnan10}, and compute the uncorrectable bit error rate using
the formulas in \cite{Mielke08}.  Then as described in Section~\ref{subsec:
parameter}, we derive $c$ for different ECCs that can correct 3, 4, 5 bits
per 512 byte sector, and the corresponding values are $4.4\times10^{-11}$,
$4.7\times10^{-14}$, and $4.2\times10^{-17}$, respectively. We observe that
the RAID reliability drops to zero very quickly for 3-bit ECC at around
$10^5$ erasures, while the RAID reliability  for 5-bit ECC starts to
decrease until the array performs $10^{11}$ erasures. This shows that the
RAID reliability heavily depends on the reliability of each single SSD, or
the ECC length employed in each SSD.

\noindent {\bf Impact of $M$.} Figure~\ref{fig: array_conf}(c) shows the
impact of the erasure limit $M$, or the endurance of a single SSD, on the
RAID reliability. We fix other parameters with $\mu=10^{-3}$, $N=9$ and
$c=0.4\times10^{-13}$. We observe that when $M$ decreases, the RAID
reliability increases.  For example, at $1.3\times 10^{10}$ erasures, the
RAID reliability increases from 0.85 to 0.99 when $M$ decreases from 10K to
1K.  Recall that the error rates increase with the number of erasures in
SSDs. We now have the increase of bit error rates capped by the small
erasure limit. The trade-off is that the SSDs are worn out and replaced more
frequently with smaller $M$.

\subsection{Discussion}

Our results provide several insights into constructing RAID for SSDs.
\begin{itemize}
\item The error dominant case may correspond to the low-end MLC or TLC
    SSDs with high bit error rates, especially when these types of SSDs
    have low I/O bandwidth for RAID reconstruction.  Both traditional
    RAID-5 and Diff-RAID show low reliability.  A higher degree of fault
    tolerance (e.g., using RAID-6 or stronger ECC) becomes necessary in
    this case.
\item When the error arrival and recovery rates are similar, Diff-RAID,
    with uneven parity distribution, achieves higher reliability than
    RAID-5, especially when RAID-5 reaches zero reliability when all SSDs
    are worn out simultaneously. This conforms to the findings in
    \cite{Balakrishnan10}.
\item In the recovery dominant case, which may correspond to the high-end
    single-level cell (SLC) SSDs that typically have very small bit error
    rates, RAID-5 achieves very high reliability. We may choose RAID-5
    over Diff-RAID in RAID deployment to save the overhead of parity
    redistribution in Diff-RAID.
\item Our model can effectively analyze the RAID reliability with regard
    to different RAID configurations.
\end{itemize}

\section{{\bf Related Work}}
\label{sec: related}

There have been extensive studies on NAND flash-based SSDs. A detailed
survey  of the algorithms and data structures for flash memories is found in
\cite{Gal05}.  Recent papers empirically study the  intrinsic
characteristics of SSDs (e.g., \cite{Agrawal08,Chen09}), or develop
analytical models for the write performance (e.g., \cite{Desnoyers12,Hu09})
and garbage collection algorithms (e.g., \cite{Li13}) of SSDs.

Bit error rates of SSDs are known to increase with the number of erase
cycles \cite{Mielke08,Grupp09}. To improve reliability, prior studies
propose to adopt RAID for SSDs at the device level \cite{Balakrishnan10,
Im11, Lee11, Lee09, Park09, Mao12}, or at the chip level \cite{Kim12}.
These studies focus on developing new RAID schemes that improve the
performance and endurance of SSDs over traditional RAID.  The performance
and reliability implications of RAID on SSDs are also experimentally studied
in \cite{Jeremic11}.  In contrast, our work focuses on quantifying
reliability dynamics of SSD RAID from a theoretical perspective.  Authors of
Diff-RAID \cite{Balakrishnan10} also attempt to quantify the reliability,
but they only compute the reliability at the instants of SSD replacements,
while our model captures the time-varying nature of error rates in SSDs and
quantifies the instantaneous reliability during the whole lifespan of an SSD
RAID array.

RAID was first introduced in \cite{Patterson88} and has been widely used in
many storage systems.  Performance and reliability analysis on RAID in the
context of hard disk drives has been extensively studied (e.g., see
\cite{Muntz90,Chen96,Burkhard93,Malhotra93,Wu97}).  On the other hand, SSDs
have a distinct property that their error rates increase as they wear down,
so a new model is necessary to characterize the reliability of SSD RAID.

\section{{\bf Conclusions}}
\label{sec: conclusion}

We develop the {\em first} analytical model that quantifies the reliability
dynamics of SSD RAID arrays.  We build our model as a non-homogeneous
continuous time Markov chain, and use uniformization to analyze the
transient state of the RAID reliability.  We validate the correctness of our
model via trace-driven DiskSim simulation with SSD extensions.

One major application of our model is to characterize the reliability
dynamics of general RAID schemes with different parity placement
distributions.  To demonstrate, we compare the reliability dynamics of the
traditional RAID-5 scheme and the new Diff-RAID scheme under different error
patterns and different array configurations.  Our model provides a useful
tool for system designers to understand the reliability of an SSD RAID array
with regard to different scenarios.

\bibliographystyle{abbrv}
\bibliography{ref}

\begin{thebibliography}{10}

\bibitem{Agrawal08}
N.~Agrawal, V.~Prabhakaran, T.~Wobber, J.~D. Davis, M.~Manasse, and
  R.~Panigrahy.
\newblock {Design Tradeoffs for SSD Performance}.
\newblock In {\em Proc. of USENIX ATC}, Jun 2008.

\bibitem{Balakrishnan10}
M.~Balakrishnan, A.~Kadav, V.~Prabhakaran, and D.~Malkhi.
\newblock {Differential RAID: Rethinking RAID for SSD Reliability}.
\newblock {\em ACM Trans. on Storage}, 6(2):4, Jul 2010.

\bibitem{disksim}
J.~S. Bucy, J.~Schindler, S.~W. Schlosser, and G.~R. Ganger.
\newblock {The DiskSim Simulation Environment Version 4.0 Reference Manual}.
\newblock Technical Report CMUPDL-08-101, May 2008.

\bibitem{Burkhard93}
W.~Burkhard and J.~Menon.
\newblock {Disk Array Storage System Reliability}.
\newblock In {\em Proc. of IEEE FTCS}, Jun 1993.

\bibitem{Chen09}
F.~Chen, D.~A. Koufaty, and X.~Zhang.
\newblock {Understanding Intrinsic Characteristics and System Implications of
  Flash Memory Based Solid State Drives}.
\newblock In {\em SIGMETRICS}, 2009.

\bibitem{Chen96}
S.~Chen and D.~Towsley.
\newblock {A Performance Evaluation of RAID Architectures}.
\newblock {\em IEEE T. on Comp.}, 45(10):1116--1130, 1996.

\bibitem{de00}
E.~de~Souza~e Silva and H.~R. Gail.
\newblock {Transient Solutions for Markov Chains}.
\newblock {\em Computational Probability}, W. K. Grassmann (editor). Kluwer
  Academic Publishers:43--81, 2000.

\bibitem{Deal09}
E.~Deal.
\newblock {Trends in NAND Flash Memory Error Correction}.
\newblock
  \url{http://www.cyclicdesign.com/whitepapers/Cyclic_Design_NAND_ECC.pdf}, Jun
  2009.

\bibitem{Desnoyers12}
P.~Desnoyers.
\newblock {Analytic Modeling of SSD Write Performance}.
\newblock In {\em Proc. of SYSTOR}, Jun 2012.

\bibitem{SSDinDataCenter2}
R.~Enderle.
\newblock {Revolution in January: EMC Brings Flash Drives into the Data
  Center}.
\newblock \url{http://www.itbusinessedge.com/blogs/rob/?p=184}, Jan 2008.

\bibitem{Gal05}
E.~Gal and S.~Toledo.
\newblock {Algorithms and Data Structures for Flash Memories}.
\newblock {\em ACM Comput. Surv.}, 37(2):138--163, 2005.

\bibitem{Grupp09}
L.~M. Grupp, A.~M. Caulfield, J.~Coburn, S.~Swanson, E.~Yaakobi, P.~H. Siegel,
  and J.~K. Wolf.
\newblock {Characterizing Flash Memory: Anomalies, Observations, and
  Applications}.
\newblock In {\em Proc. of IEEE/ACM MICRO}, Dec 2009.

\bibitem{Grupp12}
L.~M. Grupp, J.~D. Davis, and S.~Swanson.
\newblock {The Bleak Future of NAND Flash Memory}.
\newblock In {\em USENIX FAST}, Feb 2012.

\bibitem{SSDinDataCenter3}
K.~Hess.
\newblock {2011: Year of the SSD?}
\newblock
  \url{http://www.datacenterknowledge.com/archives/2011/02/17/2011-year-of-the-ssd/},
  Feb 2011.

\bibitem{Hu09}
X.-Y. Hu, E.~Eleftheriou, R.~Haas, I.~Iliadis, and R.~Pletka.
\newblock {Write Amplification Analysis in Flash-based Solid State Drives}.
\newblock In {\em Proc. of SYSTOR}, May 2009.

\bibitem{Im11}
S.~Im and D.~Shin.
\newblock {Flash-Aware RAID Techniques for Dependable and High-Performance
  Flash Memory SSD}.
\newblock {\em IEEE Trans. on Computers}, 60:80--92, Jan 2011.

\bibitem{intelspec}
Intel.
\newblock {Intel Solid-State Drive 710: Endurance. Performance. Protection}.
\newblock
  \url{http://www.intel.com/content/www/us/en/solid-state-drives/solid-state-drives-710-series.html}.

\bibitem{Jensen53}
A.~Jensen.
\newblock {Markoff Chains As An Aid in The Study of Markoff Processes}.
\newblock {\em Scandinavian Actuarial Journal}, 3:87--91, 1953.

\bibitem{Jeremic11}
N.~Jeremic, G.~M\"{u}hl, A.~Busse, and J.~Richling.
\newblock {The Pitfalls of Deploying Solid-state Drive RAIDs}.
\newblock In {\em SYSTOR}, 2011.

\bibitem{Kim12}
J.~Kim, J.~Lee, J.~Choi, D.~Lee, and S.~H. Noh.
\newblock {Enhancing SSD Reliability Through Efficient RAID Support}.
\newblock In {\em Proc. of APSys}, Jul 2012.

\bibitem{Lee11}
S.~Lee, B.~Lee, K.~Koh, and H.~Bahn.
\newblock {A Lifespan-aware Reliability Scheme for RAID-based Flash Storage}.
\newblock In {\em Proc. of ACM Symp. on Applied Computing}, SAC '11, 2011.

\bibitem{Lee09}
Y.~Lee, S.~Jung, and Y.~H. Song.
\newblock {FRA: A Flash-aware Redundancy Array of Flash Storage Devices}.
\newblock In {\em Proc. of ACM CODES+ISSS}, Oct 2009.

\bibitem{Li13}
Y.~Li, P.~P.~C. Lee, and J.~C.~S. Lui.
\newblock {Stochastic Modeling of Large-Scale Solid-State Storage Systems:
  Analysis, Design Tradeoffs and Optimization}.
\newblock In {\em Proc. of SIGMETRICS}, 2013.

\bibitem{Malhotra93}
M.~Malhotra and K.~S. Trivedi.
\newblock {Reliability Analysis of Redundant Arrays of Inexpensive Disks}.
\newblock {\em J. Parallel Distrib. Comput.}, 17(1-2):146--151, Jan 1993.

\bibitem{Mao12}
B.~Mao, H.~Jiang, S.~Wu, L.~Tian, D.~Feng, J.~Chen, and L.~Zeng.
\newblock {HPDA: A Hybrid Parity-based Disk Array for Enhanced Performance and
  Reliability}.
\newblock {\em ACM Trans. on Storage}, 8(1):4, Feb 2012.

\bibitem{Mariano12}
M.~Mariano.
\newblock {ECC Options for Improving NAND Flash Memory Reliability}.
\newblock
  \url{http://www.micron.com/~/media/Documents/Products/Software%20Article/SWNL_implementing_ecc.pdf},
  Nov 2011.

\bibitem{Mielke08}
N.~Mielke, T.~Marquart, N.~Wu, J.~Kessenich, H.~Belgal, E.~Schares, F.~Trivedi,
  E.~Goodness, and L.~Nevill.
\newblock {Bit Error Rate in NAND Flash Memories}.
\newblock In {\em IEEE Int. Reliability Physics Symp.}, Apr 2008.

\bibitem{Muntz90}
R.~R. Muntz and J.~C.~S. Lui.
\newblock {Performance Analysis of Disk Arrays under Failure}.
\newblock In {\em Proc. of VLDB}, Aug 1990.

\bibitem{Narayanan09}
D.~Narayanan, E.~Thereska, A.~Donnelly, S.~Elnikety, and A.~Rowstron.
\newblock {Migrating Server Storage to SSDs: Analysis of Tradeoffs}.
\newblock In {\em Proc. of ACM EuroSys}, Mar 2009.

\bibitem{Park09}
K.~Park, D.-H. Lee, Y.~Woo, G.~Lee, J.-H. Lee, and D.-H. Kim.
\newblock {Reliability and Performance Enhancement Technique for SSD Array
  Storage System Using RAID Mechanism}.
\newblock In {\em IEEE Int. Symp. on Comm. and Inform. Tech.}, 2009.

\bibitem{Patterson88}
D.~A. Patterson, G.~Gibson, and R.~H. Katz.
\newblock {A Case for Redundant Arrays of Inexpensive Disks (RAID)}.
\newblock In {\em Proc. of ACM SIGMOD}, Jun 1988.

\bibitem{Reibman89}
A.~Reibman and K.~S. Trivedi.
\newblock {Transient Analysis of Cumulative Measures of Markov Model Behavior}.
\newblock {\em Communications in Statistics-Stochastic Models}, 5:683--710,
  1989.

\bibitem{Schulze89}
M.~Schulze, G.~Gibson, R.~Katz, and D.~Patterson.
\newblock {How Reliable Is A RAID?}
\newblock In {\em IEEE Computer Society International Conference: Intellectual
  Leverage, Digest of Papers}, 1989.

\bibitem{Weibull51}
W.~Weibull.
\newblock {A Statistical Distribution Function of Wide Applicability}.
\newblock {\em J. of Applied Mechanics}, 18:293--297, 1951.

\bibitem{Wu97}
X.~Wu, J.~Li, and H.~Kameda.
\newblock {Reliability Analysis of Disk Array Organizations by Considering
  Uncorrectable Bit Errors}.
\newblock In {\em Proc. of IEEE SRDS}, Oct 1997.

\end{thebibliography}

\appendix

\subsection{ Proof of Theorem \ref{theo: matrix_perturbation} in Section
\ref{subsec: analyze_nonhomogeneous} }

The computation of the system state in Equation (\ref{eq:
system_state_mp_u_t}) is intuitive since the truncation point is $U_l$ in
interval $(lsT, (l+1)sT)$. In the following, we focus on the derivation of
the error bound. Note that $\bpi((l+1)sT)$ is the system state at time
$(l+1)sT$ for the CTMC $\{X(t)\}$. Moreover, given the state at time $lsT$,
$\bpi((l+1)sT)$ is computed iteratively by  computing $\bpi((ls+1)T)$,
$\bpi((ls+2)T)$, ..., $\bpi((ls+s)T)$ sequentially. During each step, e.g.,
deriving $\bpi((k+1)T)$ from $\bpi(kT)$ ($ls\leq k < (l+1)s$),
uniformization is used. Without loss of generality, we can let
$\Lambda_k=\tilde{\Lambda}_l$ ($ls\leq k < (l+1)s$) as
$\tilde{\Lambda}_l\geq\max_{0\leq i\leq S+1}|-q_{i,i}(k)|$ for all $k$
($ls\leq k < (l+1)s$). Since $\bQ_k$ is denoted as the generator matrix of
the homogeneous CTMC $\{X(t), kT<t\leq (k+1)T\}$, to apply the
uniformization, we let  $\bP_k = \bI+\frac{\bQ_k}{\tilde{\Lambda}_l}$
($ls\leq k < (l+1)s$). Since every element of $\bP_k$ is a linear function
of $k$, the difference between two matrices $\bP_{k+1}-\bP_k$ must be the
same for all $k$, and we denote it by $\bD$. Formally, we have
\begin{equation}
  \bD=\bP_{k+1}-\bP_k, \; ls\leq k < (l+1)s
\label{eq: matrixd}
\end{equation}
Now, we can easily find that $\bP_k=\bP_{ls}+(k-ls)\bD$ ($ls \leq k \leq
(l+1)s-1$). Moreover, since  $\tilde{\bP_l} = \bI +
\frac{\tilde{\bQ}_l}{\tilde{\Lambda}_l}$ and  $\tilde{\bQ}_l$ is defined as
$\frac{\sum_{k=ls}^{(l+1)s-1}\bQ_k}{s}$ in Equation (\ref{eq: new_generator
matrix_avg}), we have
\begin{eqnarray}
  \tilde{\bP_l}
 &=&\frac{\sum_{k=ls}^{(l+1)s-1}\bP_k}{s}=\frac{\sum_{k=ls}^{(l+1)s-1}(\bP_{ls}+(k-ls)\bD)}{s}\nonumber\\
&=&\bP_{ls}+\frac{s-1}{2}\bD.
\label{eq: new_p_avg}
\end{eqnarray}

Note that based on the analysis of $\{X(t), kT<t\leq (k+1)T\}$ by using
uniformization, $\bpi((k+1)T)$ ($ls \leq k < (l+1)s$) can be rewritten as
follows.
\begin{eqnarray*}
  \bpi((k+1)T)&=&\bpi(kT)e^{-\tilde{\Lambda}_l T}e^{\tilde{\Lambda}_l T \bP_k}\\
&=&\bpi(kT)e^{-\tilde{\Lambda}_l T}e^{\tilde{\Lambda}_l T (\bP_{ls}+(k-ls)\bD)}.
\end{eqnarray*}
Observe that most elements in the difference matrix $\bD$ are zero, and the
non-zero elements are all very small, by examining the elements in
$\bD\bP_{ls}$ and the elements in $\bP_{ls}\bD$,  we find that the
multiplication of matrix $\bD$ and matrix $\bP_{ls}$ can be assumed to be
commutative, or $\bD\bP_{ls}\approx\bP_{ls}\bD$. Therefore, we have
\begin{eqnarray*}
  \bpi((l+1)sT)&\approx&\bpi(lsT)e^{-\tilde{\Lambda}_l sT}e^{\tilde{\Lambda}_l T \sum_{k=ls}^{(l+1)s-1} \bP_k}\\
&=&\bpi(lsT)e^{-\tilde{\Lambda}_l sT}e^{\tilde{\Lambda}_l T s\tilde{\bP}_l}.
\end{eqnarray*}

Now, the upper bound of the error $\hat{\tilde{\epsilon}}_l$ is derived as
follows.
\begin{eqnarray*}
\hat{\tilde{\epsilon}}_l\!\!\!\!\!\!&=&\!\!\!\!\!\!||\hat{\tilde{\bpi}}((l+1)sT) -\bpi((l+1)sT)||_1\\
\!\!\!\!\!\!&=&\!\!\!\!\!\!||\hat{\tilde{\bpi}}(lsT)\!\!\!\sum_{n=0}^{U_l}\!\!e^{\!-\!\tilde{\Lambda}_l s T}\!\frac{(\tilde{\Lambda}_l sT)^n}{n!}\!\tilde{\bP}_l^n
\!\!\!-\!\!\bpi(lsT)e^{\!-\!\tilde{\Lambda}_l sT}e^{\tilde{\Lambda}_l T s\tilde{\bP}_l}\!||_1\\
\!\!\!\!\!\!&\leq&\!\!\!\!\!\!||\hat{\tilde{\bpi}}(lsT)e^{-\tilde{\Lambda}_l s T}e^{\tilde{\Lambda}_l s T\tilde{\bP}_l}
\!-\!\bpi(lsT)e^{-\tilde{\Lambda}_l sT}e^{\tilde{\Lambda}_l T s\tilde{\bP}_l}||_1\\
&&+ ||\hat{\tilde{\bpi}}(lsT)\!\!\sum_{n=U_l+1}^{\infty}\!\!e^{-\tilde{\Lambda}_l s T}\frac{(\tilde{\Lambda}_l sT)^n}{n!}\tilde{\bP}_l^n||_1\\
\!\!\!\!\!\!&\leq&\!\!\!\!\!\!||\hat{\tilde{\bpi}}(lsT)-\bpi(lsT)||_1e^{-\tilde{\Lambda}_l s T}e^{\tilde{\Lambda}_l s T||\tilde{\bP}_l||_{\infty}}\\
&&+ \left(1-\sum_{n=0}^{U_l}e^{-\tilde{\Lambda}_l s T}\frac{(\tilde{\Lambda}_l sT)^n}{n!}\right)\\
\!\!\!\!\!\!&=&\!\!\!\!\!\!\hat{\tilde{\epsilon}}_{l-1}+\left(1-\sum_{n=0}^{U_l}e^{-\tilde{\Lambda}_l s T}\frac{(\tilde{\Lambda}_l sT)^n}{n!}\right).
\end{eqnarray*}
The  last equation comes from the fact that $||\tilde{\bP}_l||_{\infty}=1$
as $\tilde{\bP}_l=\bI+\frac{\tilde{\bQ}_l}{\tilde{\Lambda}_l}$, and
$\hat{\tilde{\epsilon}}_{l-1}=||\hat{\tilde{\bpi}}(lsT) -\bpi(lsT)||_1$.
Therefore, we have the results stated in Theorem \ref{theo:
matrix_perturbation}. \done

\end{document}